\documentclass[fleqn,usenatbib]{mnras}

\usepackage{newtxtext,newtxmath}

\usepackage{multirow}

\usepackage[T1]{fontenc}
\usepackage{natbib}

\DeclareRobustCommand{\VAN}[3]{#2}
\let\VANthebibliography\thebibliography
\def\thebibliography{\DeclareRobustCommand{\VAN}[3]{##3}\VANthebibliography}

\usepackage{graphicx}	
\usepackage{amsmath}	
\usepackage{ulem}
\usepackage{amssymb}	
\usepackage{hyperref}

\hypersetup{
    plainpages=false,       
    unicode=false,          
    pdftoolbar=true,        
    pdfmenubar=true,        
    pdffitwindow=false,     
    pdfstartview={FitH},    
    pdfnewwindow=true,      
    colorlinks=true,        
    linkcolor=blue,         
    citecolor=blue,        
    filecolor=magenta,      
    urlcolor=cyan           
}

\title[Positive radio-mechanical feedback in A1795]{Radio jet-ISM interaction and positive radio-mechanical feedback in Abell 1795}

\author[Tamhane et al.]{
Prathamesh D. Tamhane,$^{1,2}$\thanks{E-mail: pdtamhane@uwaterloo.ca}
Brian R. McNamara,$^{1,2,3}$
Helen R. Russell,$^{4}$
Francoise Combes,$^{5}$
\newauthor Yu Qiu,$^{6}$
Alastair C. Edge,$^{7}$
Roberto Maiolino,$^{8}$
Andrew C. Fabian,$^{9}$
\newauthor Paul E.J. Nulsen,$^{10,11}$
R. Johnstone$^{9}$
and Stefano Carniani$^{12}$
\\
$^{1}$Waterloo Centre for Astrophysics, University of Waterloo, Waterloo, ON, N2L 3G1, Canada\\
$^{2}$Department of Physics and Astronomy, University of Waterloo, Waterloo, ON, N2L 3G1, Canada\\
$^{3}$Perimeter Institute for Theoretical Physics, Waterloo, ON, N2L 2Y5, Canada\\
$^{4}$Department of Physics and Astronomy, University of Nottingham, University Park, Nottingham NG7 2RD, UK\\
$^{5}$LERMA, Observatoire de Paris, Coll\`ege de France, PSL Univ., CNRS, Sorbonne Univ., 75014, Paris France\\
$^{6}$Kavli Institute for Astronomy and Astrophysics, Peking University, 5 Yiheyuan Road, Haidian District, Beijing, 100871, People’s Republic of China\\
$^{7}$Department of Physics, Durham University, Durham DH1 3LE, UK\\
$^{8}$Cavendish Laboratory, Department of Physics, University of Cambridge, UK\\
$^{9}$Institute of Astronomy, Madingley Road, Cambridge CB3 0HA, UK\\
$^{10}$Harvard-Smithsonian Center for Astrophysics, 60 Garden Street, Cambridge, MA 02138, USA\\
$^{11}$ICRAR, University of Western Australia, 35 Stirling Hwy, Crawley, WA 6009, Australia\\
$^{12}$Scuola Normale Superiore, Piazza dei Cavalieri 7, I-56126 Pisa, Italy\\
}

\date{Accepted XXX. Received YYY; in original form ZZZ}

\pubyear{2022}

\begin{document}
\label{firstpage}
\pagerange{\pageref{firstpage}--\pageref{lastpage}}
\maketitle

\begin{abstract}
We present XSHOOTER observations with previous ALMA, MUSE and $HST$ observations to study the nature of radio-jet triggered star formation and the interaction of radio jets with the interstellar medium in the brightest cluster galaxy (BCG) in the Abell 1795 cluster. Using $HST$ UV data we determined an ongoing star formation rate of 9.3 M$_\odot$ yr$^{-1}$. The star formation follows the global Kennicutt-Schmidt law, however, it has a low efficiency compared to circumnuclear starbursts in nearby galaxies with an average depletion time of $\sim$1 Gyr. The star formation and molecular gas are offset by $\sim1$ kpc indicating that stars have decoupled from the gas. We detected an arc of high linewidth in ionized gas where electron densities are elevated by a factor of $\sim$4 suggesting a shock front driven by radio jets or peculiar motion of the BCG. An analysis of nebular emission line flux ratios suggests that the gas is predominantly ionized by star formation with a small contribution from shocks. We also calculated the velocity structure function (VSF) of the ionized and molecular gases using velocity maps to characterize turbulent motion in the gas. The ionized gas VSF suggests that the radio jets are driving supersonic turbulence in the gas. Thus radio jets can not only heat the atmosphere on large scales and may quench star formation on longer timescales while triggering star formation in positive feedback on short timescales of a few million years.
\end{abstract}




\section{Introduction}
 \label{sec:intro}

The Active Galactic Nucleus (AGN) feedback has become an integral part of galaxy simulations to limit the growth of massive galaxies via star formation and to explain the presence of young stellar populations in low redshift galaxies \citep[eg.][]{kauffman00,schawinski06,croton06,scannapieco12}. Recent discoveries of massive gas outflows driven by intense radiation or radio-jets have demonstrated how effectively the energy released by an AGN can interact with its host environment \citep[eg.][]{morganti05,nesvadba06,feruglio10,ruffa19,ruffa22}. By heating and expelling cold gas from a galaxy’s centre, the AGN activity can limit the fuel available for star formation and accretion onto the supermassive black hole (SMBH) and thereby regulate galaxy growth.

The AGN feedback mainly occur in two phases, the ``quasar mode'' and the ``radio mode''. The quasar mode occurs when the cold gas accretes efficiently on the central SMBH, forming an accretion disk that emits at all wavelengths. The intense radiation and winds from the accretion disk disrupt the surrounding gas and terminate star formation in a negative feedback \citep{fabian12,tombesi15,veilleux20}. In the low accretion efficiency regime, the radio mode feedback dominates \citep{russell13}. The radio jets inject energy into the atmospheres of galaxies and galaxy clusters by blowing bubbles, often seen as cavities or surface brightness depressions in X-ray images. The atmosphere is heated by a combination of shocks \citep{randall11}, turbulence \citep{zhuravleva14}, and the enthalpy released as the cavities rise, leading to suppression of cooling and star formation \citep{mcnamara07,mcnamara12,donahue22}.

Although radio mode feedback is thought to maintain balance in the inner hot atmospheres of galaxies and clusters, AGN with pronounced radio jets are also prone to positive feedback \citep{zinn13,gaibler12}. Theoretical models predict positive feedback
scenarios where a collimated jet or outflow compresses the surrounding interstellar medium (ISM) to trigger star formation, and stars form
within radiatively-driven outflows as this material is compressed, cools and fragments \citep[eg.][]{silk13,ishibashi12,zubovas13,bieri16}. Star formation rates from jet-ISM interactions are observed to be
limited \citep[eg.][]{vanb85,crockett12,cresci15,salome16,santoro16}. However, a different class of models and numerical hydrodynamical simulations have predicted an even more intriguing scenario where stars can form {\it inside} the outflowing clouds \citep{ishibashi12,fabian12,zubovas13,yu20}. Some of these models expect that stars can form in this mode at a rate of up to 10-100 M$_\odot$ yr$^{-1}$.
This exciting new mode of star formation
would generate stars with high radial velocities, hence with kinematic properties different with respect to the galactic disk, and could contribute to the formation of the spheroidal components of galaxies (bulge, halo). Therefore, star
formation in these flows would contribute to the morphological evolution of galaxies \citep[eg.][]{gaibler12}
and could explain key aspects of galactic structure, including the mass-radius relation of early-type galaxies
\citep[eg.][]{ishibashi14} and `inside-out' galaxy growth, where compact $z\sim2$ galaxies form the cores
of present-day massive ellipticals \citep{vandokkum10,ishibashi12}.

Radio-jet induced star formation within their host galaxies has been detected in many objects at low \citep{nesvadba21,salome17,salome15} and high redshifts \citep{emonts14, nesvadba20}. 
Recent Very Large Telescope (VLT) observations of local ultra-luminous infrared galaxy (ULIRG) IRAS F23128-5919 obtained the first detection of star formation in
outflowing gas occurring at a rate of 30 $M_\odot$ yr$^{-1}$, which is $\sim$ 25 percent of the global star formation rate in this system \citep{maiolino17}. Although with lower resolution, recent studies using the MaNGA survey have found evidence for positive feedback and star formation in outflows in a significant fraction of galaxies \citep{gallagher19,rodriguez19}. Here we study another example of well-resolved star formation within molecular gas flows driven by radio jets in the BCG of Abell 1795 galaxy cluster. Its proximity and deep, high-resolution multiwavelength observations make it an excellent candidates to study the positive radio-mechanical feedback.

The central elliptical galaxy of the nearby cluster Abell 1795 hosts a powerful radio source, 4C 26.42, which is driving two massive 10$^9$ $M_\odot$ molecular gas flows \citep[Fig. 3][]{russell17b}. These extended cool gas filaments are each $\sim$ 7 kpc in length and visible in soft
X-ray, H$\alpha$, CO and bright streams of star formation \citep[eg.][]{mcnamara93,mcnamara96,salome04,crawford05}.
The filaments project exclusively around the outer edges of two radio bubbles that have been inflated by the jet \citep{vanb84,fabian01}. The close spatial association with the radio lobes, together with smooth velocity gradients along their lengths and narrow velocity
dispersions, show that the molecular filaments are gas flows entrained by the expanding bubbles as explained in \citep{russell17b}. For thick,
clumpy shells of molecular gas around the radio bubbles, the column depth is greatest around the peripheries,
and therefore this morphology will be detected as bright rims, as observed.
The south (S) molecular filament appears to form an extension of a bright stream of young stars that are
visible in $HST$ FUV observations, but appears disconnected from the central clump and spatially offset from the UV emission (shown in Fig.~\ref{fig:ks_new}). This striking spatial anti-correlation suggests that the molecular gas flow
could be fuelling this burst of star formation. In this paper, we study the interplay between the radio-jets, ionized and molecular gas and star formation using high spatial resolution multiwavelength observations of Abell 1795.

Throughout this paper, we use flat $\Lambda$CDM cosmology with $H_0$ = 70 km s$^{-1}$ Mpc$^{-1}$, $\Omega_{\rm m}$ = 0.3 and $\Omega_{\Lambda}$ = 0.7. We adopt a redshift of 0.063001 (see section~\ref{redshift}), where 1 arcsec corresponds to 1.21 kpc. This paper is structured as follows: section~\ref{sec:data} describes the data used in this paper, section~\ref{sec:analysis} describes stellar kinematics, ionized and molecular gas morphology and kinematics, section~\ref{COHacomp} discusses the comparison of ionized and molecular gas morphologies and velocities, section~\ref{sf} discusses star formation and its efficiency, section~\ref{elineratios} discusses various nebular emission line ratios in the BCG, section~\ref{vsf} describes the velocity structure function of the ionized and molecular gases and its discussion and finally we summarize our results in section~\ref{conclusion}.

\section{Observations and Data Reduction} \label{sec:data}
\subsection{XSHOOTER}
\label{xshoo}
\begin{figure*}
		\centering
		\includegraphics[width=\textwidth]{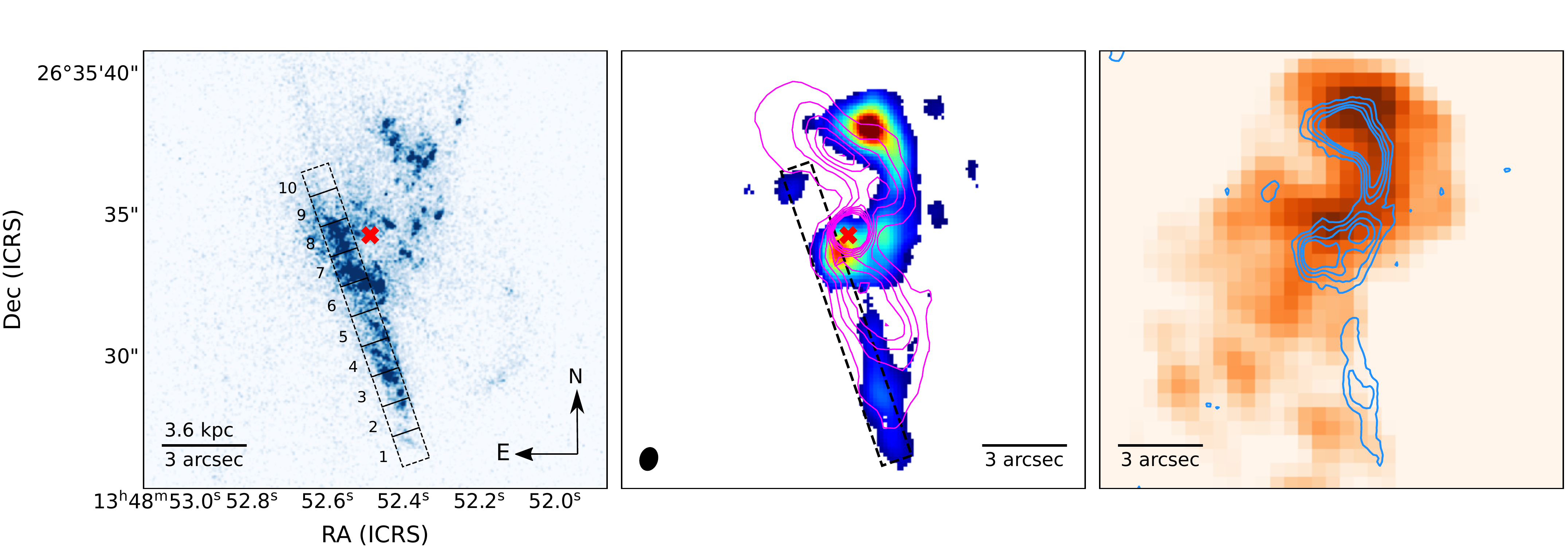}
		\caption{Left: $HST$F150LP image of the star formation in Abell 1795. The black dashed box shows the position of the X-SHOOTER slit. The numbers indicate the regions along the slit used to extract the spectra for studying gas and stellar kinematics. Center: Integrated CO(2-1) total intensity map with ALMA for velocities between $-$340 and 150 km s$^{-1}$. The dashed black box represents the orientation of the XSHOOTER slit relative to the molecular gas. We show radio contours at 8 GHz from the Very Large Array (VLA) from \citet{birzan08} in magenta. Right: $Chandra$ 0.5--7 keV X-ray image shows the hot X-ray gas emission with molecular gas contours from CO overlaid in blue, where levels correspond to flux densities of 0.2, 0.3, 0.4 and 0.5 Jy/beam km/s. The red ``X'' denotes the location of the nuclear radio source.}
		\label{fig:hac}
\end{figure*}

The galaxy spectra were obtained with the X-SHOOTER spectrometer at the European Southern Observatory’s
Very Large Telescope (ESO-VLT) \citep{vernet11} (Program ID: 60.A-9022(C), PI: Helen Russell). We selected 11$^{\prime\prime}$-long slit for all arms. The slit width was 1$^{\prime\prime}$ in the ultraviolet B (UVB)
arm, and 0.9$^{\prime\prime}$ in the visible (VIS) and near-infrared (NIR) arms. This setting gives a spectral resolution of 5100,
8800 and 5300 in UVB, VIS and NIR arms, respectively. The full width at half maximum (FWHM) was 0.126 nm, 0.12 nm and 0.882 nm for UVB, VIS, and NIR arms corresponding to velocity resolutions of $\sim$ 88 km s$^{-1}$, 45 km s$^{-1}$ and 151 km s$^{-1}$, respectively. The observations were taken between April 2018 and March 2019. The seeing during observations varied between
0.57$^{\prime\prime}$--1.24$^{\prime\prime}$ FWHM across exposures with a mean seeing of $\sim$0.84$^{\prime\prime}$. The slit was positioned with a position angle of
19$^\circ$ as shown in the left panel of Fig.~\ref{fig:hac}.
This orientation and slit width was chosen to enable the slit to sample both the S outflow filament and the star-forming knots to the NE of the nucleus whilst
providing the required spectral resolution. Observations were executed in the OFFSET mode and interleaved with
blank sky exposures of equal duration as that of the source, obtained at about 1 arcmin from the galaxy, for sampling
the background. For each exposure, standard star observations were taken for flux calibration.
The on-source integration time was 2 hours in the UVB and NIR bands, respectively, and 1 hour in the VIS band with a total integration time of 5 hours.

We performed data reduction and calibration in the \textsc{esoreflex} software (version 2.11.0) using the recommended X-SHOOTER pipeline (version 3.5.0) \citep[][]{freudling13} following the standard steps summarised below \citep{modigliani10}. We analyzed each individual exposure independently. The data were corrected and calibrated for detector bias and dark current, detector pixel-to-pixel sensitivity variations, wavelength calibration, instrument and detector efficiency and response and background sky subtraction. The flux calibration was performed by analyzing the standard star spectrum to determine instrument and detector response. The final data product is a wavelength corrected, flux calibrated and background subtracted 2D spectrum of the slit for each exposure.

\begin{figure*}
\includegraphics[width=\textwidth]{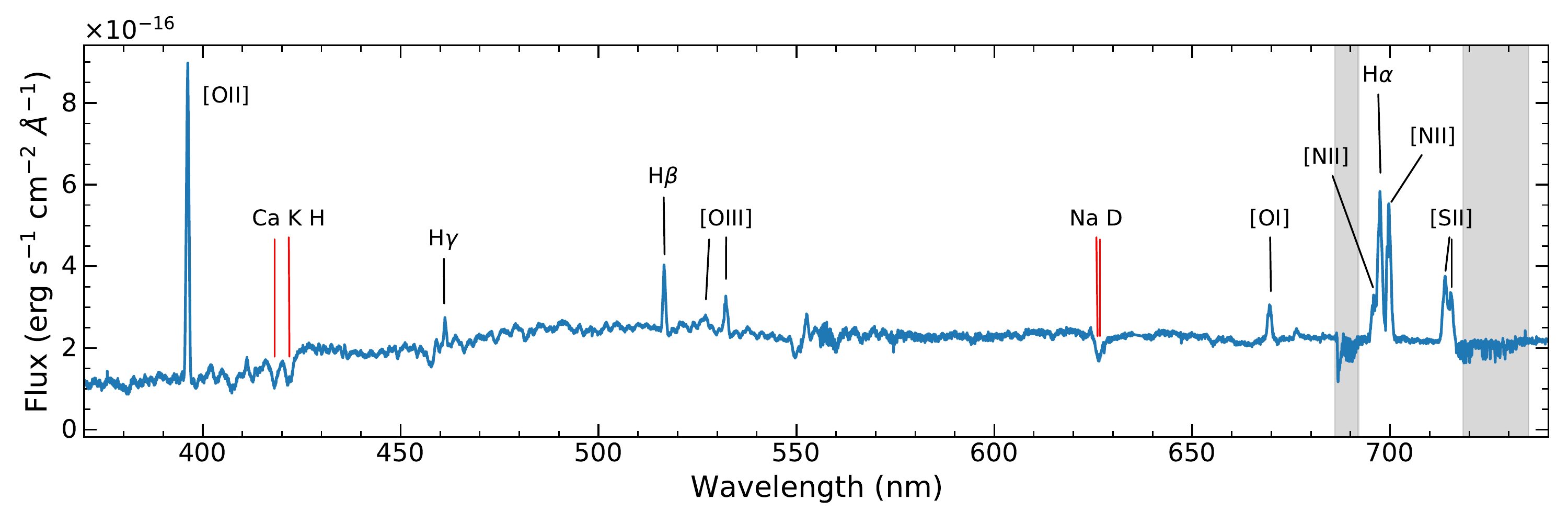}
\caption{Combined UVB and VIS arm one-dimensional XSHOOTER spectrum of the slit in the observed frame. Grey regions are affected by atmospheric telluric lines. Nebular emission lines and stellar absorption lines (red) are indicated.}
\label{fig:spec}
\end{figure*}

The observations were taken during different times of the year and the spectra generated by the pipeline were in the topocentric reference frame. Therefore, we applied radial velocity correction to change the spectrum to the barycentric frame of reference and resampled the spectrum at the original wavelength grid using \texttt{spectres}\footnote{\url{https://spectres.readthedocs.io/en/latest/}} package in Python for each exposure \citep{carnall17}. We then combined individual exposures for each arm to obtain the final 2D spectrum of the filament in UVB, VIS and NIR bands. We corrected the spectrum for Galactic foreground extinction estimated from the dust maps of \citet{schlafly11} by adopting $R_{V}$ of 3.1, the standard value for the diffuse ISM and \citet[][]{cardelli89} extinction curve. The combined UVB and VIS band spectrum is shown in Fig.~\ref{fig:spec}. We divided the area of the slit into 10 regions of 1.1$^{\prime\prime} \times 1^{\prime\prime}$ and extracted the spectrum from each region. Region 1 is $\sim$9.5 kpc to the S of the nucleus and region 10 is $\sim$3 kpc to the NE of the nucleus as shown in the left panel of Fig.~\ref{fig:hac}. The BCG nucleus lies close to region 8.

\subsection{MUSE}

We also analysed Multi-Unit Spectroscopic Explorer (MUSE) integral field unit (IFU) spectrograph data for Abell 1795. The observations were carried out on 05 February 2015 in seeing limited WFM-NOAO-N configuration (Program ID: 094.A-0859(A), PI: Hamer). The data were published and H$\alpha$ velocity maps were discussed in \citet{olivares19}. We reanalyzed the data to study other nebular emission lines together with ionized gas kinematics traced by H$\alpha$. Three exposures with an exposure time of 900s each were taken with a total on source integration time of 2700s. \citet{olivares19} analyzed only one exposure with 900 s of on source integration time. For our study, we included all exposures for better signal to noise ratio. The field of view of $1^\prime \times 1^\prime$ covered the entire BCG including extended S H$\alpha$ filament in a single pointing. Throughout the observation, the average seeing was 1.36$^{\prime\prime}$.

The data were processed with esoreflex framework using MUSE pipeline version 2.8.5 \citep{weilbacher20} with automated bias subtraction, wavelength and flux calibration, as well as illumination-, flat-field, and differential atmospheric diffraction corrections and sky subtraction. We did not perform any additional sky-subtraction. The final MUSE datacube maps the entire galaxy in the range 4750 \AA $< \lambda < $9350 \AA~with a spectral resolution of $\sim$2 \AA. The data were corrected for foreground Galactic absorption as explained in the previous section.

\subsection{ALMA}
\label{COdataAnalysis}
The Atacama Large Millimeter/submillimeter Array (ALMA) observed Abell 1795 between 11 and 14 June 2016 for 72 minutes as part of program 2015.1.00623.S (PI: Helen Russell). \citet{russell17b} analyzed the data and discussed in detail the properties of the molecular gas in Abell 1795. For the purpose of this paper, we re-analyzed the ALMA data and reproduced some of the results.

The data sets were calibrated with the ALMA pipeline reduction scripts in \textsc{
casa} version 4.7.2 \citep{mcmullin07}. We performed the standard phase calibration. Additional self-calibration did not improve the image root mean square (rms) noise. Line-free channels were used to subtract the continuum emission from the $uv$ plane using the task \textsc{uvcontsub}. We then deconvolved and imaged the continuum subtracted CO(2-1) visibilities using the \textsc{clean} algorithm with natural weighting to improve the sensitivity towards filamentary emission.

The final data cube is binned in 10 km s$^{-1}$ channels with a per channel rms sensitivity of 0.64 mJy beam$^{-1}$ and has a synthesized beam of 0.8$^{\prime\prime}$ $\times$ 0.6$^{\prime\prime}$ with a position angle of $-$15.4 deg.

We summed the flux in each pixel in the data cube in a beam-sized region centred on the pixel. Then the spectrum at each pixel was fitted with one or two Gaussian components using the {\sc lmfit} package when emission was detected at greater than the 3$\sigma$ threshold, based on Monte Carlo simulations of the spectrum with 1000 iterations. The best-fit parameters from this analysis were used to make velocity and FWHM maps of the molecular gas emission. All fluxes and linewidths are corrected for primary beam response.

\subsection{$HST$ and VLA data}
\label{hstVLAdata}
We used Hubble Space Telescope ($HST$) images taken with F150LP, F702W and F555W filters in our analysis. All images were obtained from the Hubble Legacy Archive\footnote{\url{https://hla.stsci.edu/}}. The F702W and F555W images each have total integration time of 1780 s (Proposal ID: 5212, PI: Trauger) and the F150LP image has the total integration time of 1300 s (Proposal ID: 11681, PI: Sparks).

We also used the Very Large Array (VLA) L and X band radio maps of the central radio source in Abell 1795 at 1.4 GHz and 8.2 GHz frequencies, respectively, from \citet{birzan08}.

\section{Data Analysis} \label{sec:analysis}
\subsection{Stellar kinematics}
\label{star_kin}

\begin{figure*}
    \centering
    \includegraphics[width=\textwidth]{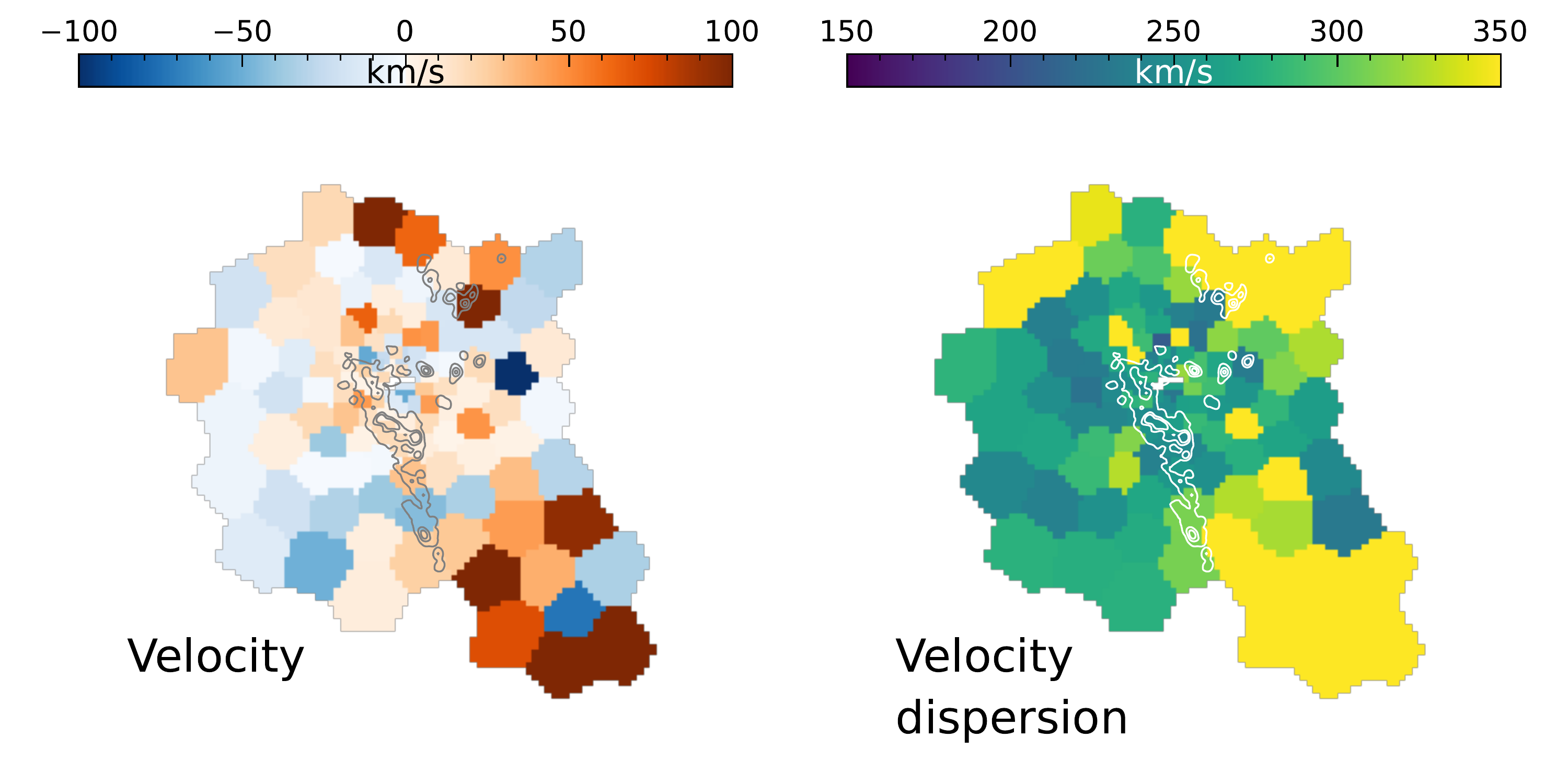}
    \caption{Voronoi binned stellar kinematics maps of A1795 with MUSE data. The left and right panels show velocity and velocity dispersion in the units of km s$^{-1}$ in each bin, respectively. Each Voronoi bin has a line-free continuum signal-to-noise ratio of at least 50. The $HST$FUV150LP image contours smoothed with a gaussian kernel of 7 pixels indicating star formation are overlaid.}
    \label{fig:stellarkin}
\end{figure*}

One of our goals is to study the motion of young O and B stars in the star forming filament of A1795 shown in the left panel of Fig.~\ref{fig:hac}. The $HST$ FUV image clearly shows the presence of young stars in the filaments as well as near the BCG centre. The spectra of young stars is dominated by He and H absorption lines at optical wavelengths. Fitting the stellar continuum containing absorption lines with young stellar population templates can provide line of sight velocities of young stars in the filament. However, stellar H absorption lines overlap with nebular gas emission lines which makes it extremely difficult to deblend them from each other. Moreover, we did not detect He absorption lines in the spectrum (see Fig.~\ref{fig:spec}). Most of the light in the spectrum is dominated by older stars in the galaxy. Therefore, it is unfeasible to determine the velocities of young stars in the S star forming filament in A1795 using current XSHOOTER data.

We estimated stellar velocities of older stars dominating the light in the optical spectrum in the BCG using the MUSE data cube. We applied the Voronoi tessellation technique using Vorbin\footnote{\url{https://pypi.org/project/vorbin/}} package \citep{cappellari03}. The MUSE data were binned into tessellations with a minimum signal-to-noise ratio of 50 in the line-free continuum. We modeled the spectrum extracted from each bin with {\sc PyParadise} package\footnote{\url{https://github.com/brandherd/PyParadise}}. {\sc PyParadise} iteratively performs linear least squares fitting of line emission masked stellar continuum spectrum of every spectral unit. It also estimates best-fit line-of-sight velocity distribution with a Markov Chain Monte Carlo (MCMC) method independently. In Figure~\ref{fig:stellarkin}, we show the stellar velocity and velocity dispersion map of the entire BCG. The stellar velocities are nearly uniform, with velocities lying between 50--100 km s$^{-1}$ of the systemic velocity. There is no indication of a velocity gradient. The stellar velocity dispersion in the central 1$^{\prime\prime}$ radius region is 259$\pm$18 km s$^{-1}$ consistent with the central stellar dispersion values reported in the literature \citep[e.g.][]{loubser18}. The stellar velocity dispersions vary between 230--320 km s$^{-1}$.

\begin{table*}
\centering
\begin{tabular}{c|ccc|ccc|}
\hline
\multirow{2}{*}{Region} & \multicolumn{3}{|c|}{Velocity (km/s)} & \multicolumn{3}{|c|}{FWHM (km/s)} \\ \cline{2-7}
 & Comp 1 & Comp 2 & Comp 3 & Comp 1 & Comp 2 & Comp 3 \\ 
\hline
1 & $-$96.66$\pm$0.25 & $-$126.26$\pm$2.50 &  & 84.61$\pm$6.39 & 225.70$\pm$6.39 &  \\ 
2 & $-$89.08$\pm$0.85 & $-$130.13$\pm$1.14 &  & 70.62$\pm$1.59 & 163.71$\pm$1.59 &  \\ 
3 & $-$185.83$\pm$0.61 & $-$114.92$\pm$1.57 &  & 57.19$\pm$1.89 & 172.03$\pm$1.89 &  \\ 
4 & $-$42.82$\pm$1.02 & $-$191.76$\pm$2.61 & 123.88$\pm$2.08 & 110.08$\pm$2.67 & 149.01$\pm$5.62 & 192.33$\pm$4.53 \\ 
5 & $-$28.76$\pm$1.03 & 109.14$\pm$1.75 & $-$154.72$\pm$9.67  & 102.37$\pm$3.00 & 185.91$\pm$3.16 & 294.02$\pm$14.92 \\ 
6 & $-$12.66$\pm$1.53 & 1.23$\pm$1.33 &  & 169.66$\pm$6.45 & 430.48$\pm$6.45 &  \\
7 & $-$31.07$\pm$1.91 & $-$364.60$\pm$2.96 & 59.89$\pm$1.70 & 129.58$\pm$5.83 & 162.59$\pm$7.58 & 482.85$\pm$3.03 \\ 
8 & $-$31.53$\pm$2.71 & $-$258.54$\pm$4.60 & 136.16$\pm$8.76 & 153.42$\pm$8.96 & 303.08$\pm$7.74 & 397.57$\pm$12.94 \\ 
9 & $-$271.38$\pm$1.16 & $-$127.74$\pm$6.91 &  & 148.38$\pm$9.26 & 576.07$\pm$9.26 &  \\ 
10 & $-$267.79$\pm$0.47 & $-$246.88$\pm$2.83 &  & 140.42$\pm$1.72 & 425.56$\pm$1.72 &  \\
\hline
\end{tabular}
\caption{The velocities and velocity FWHM of the H$\alpha$ emitting gas were measured by fitting multiple Gaussian components to their emission profiles in different regions extracted from the X-SHOOTER spectrum. The components are sorted according to their narrow to broad FWHM.}
\label{tab:gas_kin}
\end{table*}

\begin{figure*}
    \centering
    \includegraphics[width=\textwidth]{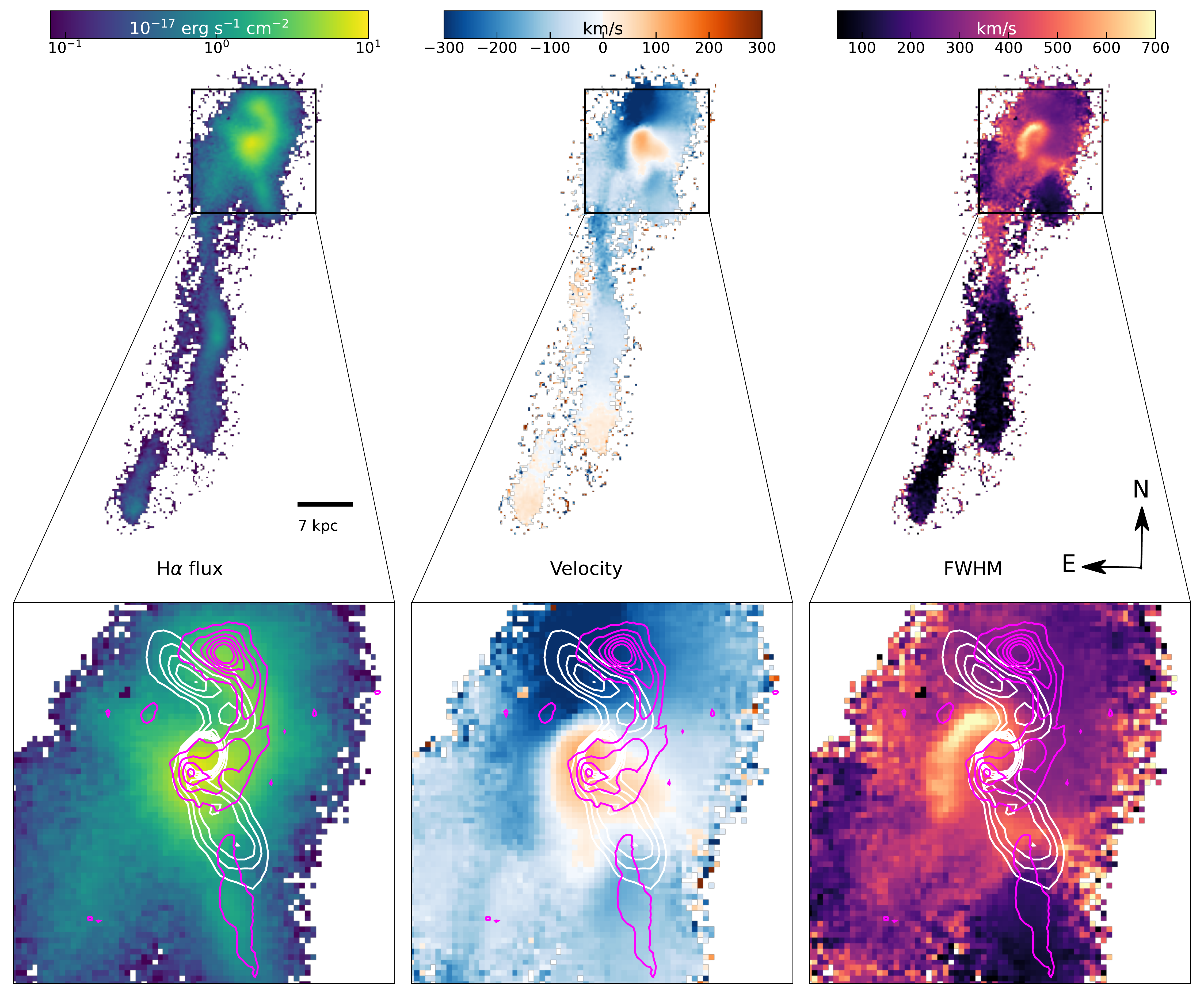}
    \caption{The figure shows the H$\alpha$ flux (left), velocity (centre) and FWHM (right) map  created from the MUSE data cube. Pixels where the FWHM is less than 800 km s$^{-1}$ are shown. The VLA radio contours at 8 GHz from \citet{birzan08} are overplotted in white. ALMA CO(2-1) contours are shown in magenta for comparison.}
    \label{fig:HaMaps}
\end{figure*}

\subsection{Systemic velocity}
\label{redshift}
We adopt a redshift of 0.063001$\pm$0.000223 corresponding to observed barycentric systemic velocity of 18887$\pm$67 km s$^{-1}$ based on stellar population synthesis modelling of the spectra extracted from central region of the BCG as described in section~\ref{star_kin}. All velocities in this paper are calculated with reference to this velocity. We note that our adopted systemic velocity is 78 km s$^{-1}$ smaller than the velocity adopted in \citet{russell17b} and \citet{olivares19}. The redshift they used is based on gas emission line, whereas we used the stellar continuum which may explain the difference between the two redshifts.

\subsection{Ionized gas mass and morphology}
\label{ionizedgas}

We used \textsc{PyParadise} to estimate the emission line fluxes in each spaxel in the MUSE data cube. The best-fit stellar continuum was generated for each spaxel as explained in the section above. It is subtracted from the spectrum, and the residuals are fit with a chain of linked Gaussians for each emission line with common velocity centroid, velocity dispersion and priors on emission line ratios. The emission lines included H$\alpha$, H$\beta$, [OIII]5007, [NII]6548,6584 and [SII]6716,6731. The errors in emission line fluxes are estimated from the Monte-Carlo bootstrap approach. The products from \textsc{PyParadise} were then used to make velocity, FWHM and flux maps of emission lines.

In Figure~\ref{fig:HaMaps} left panel, we show the flux density map of H$\alpha$ gas in the Abell 1795 BCG and the intracluster medium (ICM) around it. The peak ionized H$\alpha$ gas emission lies $\sim$0.6 kpc SE of the nuclear radio source, nearly co-spatial with the molecular gas in projection. There is a 7.6 kpc long bright filament of H$\alpha$ wrapped around the outer edge of the north (N) radio lobe. There is another filament spanning nearly 14 kpc running NE to SW of the nuclear source. The S part of this filament extending 9.5 kpc to SW of the nucleus is nearly co-spatial with the radio lobes. The N filament is twice as bright as the S filament. These filaments are surrounded by fainter H$\alpha$ emission. The MUSE FOV also covers the S outer $\sim$54 kpc long filament of H$\alpha$ that coincides with the bright filament seen in X-ray images of the cluster \citep{crawford05}.

The total H$\alpha$ luminosity in the zoomed-in region shown in Fig~\ref{fig:HaMaps} covering the BCG and inner filaments is (4.180$\pm$0.002)$\times10^{41}$ erg s$^{-1}$. We can then estimate the mass of the ionized gas as \citep{osterbrock06},
\begin{equation}
M_{\rm ion} = \frac{\mu m_H L_{\rm H_\alpha}}{\gamma n_e},
\end{equation}
where $\mu$ is the mass per hydrogen atom which we fix to 1.4, $m_H$ is the hydrogen mass, $L_{\rm H_\alpha}$ is the H$\alpha$ luminosity and $n_e$ is the electron density, which is 100 cm$^{-3}$ (see section~\ref{edensity}) and $\gamma$ is the effective line emissivity. Assuming that the ionized gas is at a temperature of 12,000 K, $\gamma \sim 3 \times 10^{-25} ~\rm erg~cm^3~s^{-1}$ \citep{baron19}, the ionized gas mass in the galaxy is (1.640$\pm$0.001)$\times 10^7~\rm M_\odot$.

\subsection{Ionized gas kinematics}
\label{Ha_kin}

\subsubsection{Velocity}
The middle panel of Fig.~\ref{fig:HaMaps} shows the velocity centroid map of H$\alpha$. The H$\alpha$ gas in the centre surrounding the central radio source is redshifted to velocities of 100 km s$^{-1}$ with respect to the BCG. The redshifted gas appears wrapped around the southern radio lobe in `V'-shaped wings that extend out to $\sim$4.8 kpc on either side. They have a smooth velocity gradient from $\sim$0 km s$^{-1}$ at the edges to 100 km s$^{-1}$ at the centre of the BCG. There is a sharp change in velocities of the ionized gas at the edges of the redshifted gas from $\sim$100 km s$^{-1}$ to $-150$ km s$^{-1}$ to the E, W and S of the `V'-shaped wings. The gas immediately to the N--NE of the redshifted gas $\sim$2 kpc away from the BCG centre is blueshifted to velocities of $-320$ km s$^{-1}$, resulting in a sharp velocity change of $\sim 420$ km s$^{-1}$. The reasons for sharp changes in velocity are not clear. It is likely that we are seeing separate H$\alpha$ filaments projected along the line of sight where the redshifted gas is on the near side and blueshifted gas is on the far side. In that case, it would indicate that the redshifted gas is inflowing and the radio jets punched a hole in the gas causing the `V'-shaped velocity structure, where the gas directly along the path of the radio jet is being blown away. Alternatively, the sharp change in velocity suggests that the region is experiencing a shock either caused by radio jets or the peculiar motion of the BCG in the cluster medium. Sloshing due to the peculiar motion of the BCG may also contribute to the observed velocity structure of the ionized gas. The sharp velocity gradient of 420 km s$^{-1}$ is close to the BCG's peculiar velocity of 368 km s$^{-1}$ with respect to the cluster's average velocity \citep{zabludoff90}. Alternatively, the gas may have cooled from the surrounding hot gas through interaction with the radio lobe in its wake. However, the H$\alpha$ luminosity observed in Abell 1795 exceeds the luminosity expected if the gas were condensing from the hot ICM \citep{mcdonald12}. Therefore, it is probably a combination of both.

\subsubsection{Velocity FWHM}
The region of the sharp changes in velocities of the ionized gas is co-incident with large FWHM, as seen in the right panel of Fig.~\ref{fig:HaMaps}. In a region that appears like an arc to the N--NE of the central radio source where the velocity gradient is most extreme, the ionized gas has very high FWHM between 600--700 km s$^{-1}$. In the centre and along the inner edges of the `V'-shaped wings, the FWHM are 400--500 km s$^{-1}$. The high FWHM along the boundary of the `V'-shaped wings also drop sharply to $\sim$250 km s$^{-1}$ to the NE of the nucleus, roughly co-spatial with the N molecular gas filament. The inner S filament has a narrow FWHM of 100--150 km s$^{-1}$. The arc of very high FWHM could be a result of beam smearing, where the region of sharp velocity gradient is poorly sampled due to large seeing during observations. It can blend spectral features at different line of sight velocities leading to flattened spectrum and increased observed line of sight velocity dispersion. For example, a similar effect has been shown to greatly exacerbate the observed line of sight velocity dispersion at the centre of a galaxy for a single-component disk model \citep{davies11}. Still, the possibility of a shock cannot be completely ruled out. Higher spatial resolution spectroscopic studies of that region may provide a deeper insight. We note that \citet{crawford05} reported a rise in FWHM in the N filament. We do not detect any significant changes in FWHM in the N filament. Perhaps the single gaussian component fit does not capture the broad component properly if it is present.

\subsubsection{XSHOOTER}
We also estimated ionized gas velocities using the XSHOOTER spectrum of the inner filament. We fitted the XSHOOTER spectrum of each region in the slit with a high-resolution single stellar population templates from \citet{rg05} using the \textsc{ppxf}\footnote{\url{https://pypi.org/project/ppxf/}} package \citep{cappellari17} to determine the best-fit stellar continuum. We then subtracted the best-fit stellar continuum from the spectrum of each region to remove stellar absorption in H Balmer lines. We fitted the residuals containing nebular gas emission line fluxes outside of \textsc{ppxf}. First, we simultaneously fitted H$\alpha$, [NII] doublet and [SII] doublets (H$\alpha$-[NII]-[SII] complex) with one Gaussian component for each line. The fit was performed such that all emission lines in this complex had the same velocities and line widths. Their intensities were allowed to vary independently. The relative intensities of the two [NII]6548,6584 lines were fixed to match the relative values of their Einstein transition coefficients. If there was significant flux in the residual with one Gaussian component, we fitted one or two more Gaussian components. The velocities and velocity widths of each component were independent of each other. The parameters of the fit were used to calculate the fluxes, velocities and line widths of all emission line components. The velocities and FWHM of the H$\alpha$ lines from the XSHOOTER spectrum are shown in Table~\ref{tab:gas_kin}. We repeated this procedure separately for [OII] doublets, H$\beta$-[OIII] complex, and [OI] emission lines.

The velocity structure of the H$\alpha$ gas is complex with several velocity components in the central regions. The average H$\alpha$ velocity of all components in each region of the slit varies from $-$110 km s$^{-1}$ in region 1 to $-250$ km s$^{-1}$ in region 10. The outer regions (regions 1,2 and 3) of the filament have a narrow FWHM between 60--80 km s$^{-1}$, but require an additional broad component with an FWHM between $\sim$150--200 km s$^{-1}$. Regions 4,5,7 and 8 required three Gaussian components to account for all line emissions with velocities between $-$364 and 123 km s$^{-1}$. The velocity FWHM of these components vary from 50 to 250 km s$^{-1}$ (see Fig.~\ref{fig:Hafits}).

\subsection{Molecular gas morphology}

\label{molmorph}

\citet{russell17b} have discussed the molecular gas morphology and kinematics in detail. We briefly summarize the main results in this and the following subsection. The majority of molecular gas in Abell 1795 is in filamentary structures to the N and S of the BCG. The S filament extends out to $\sim$ 9 kpc from the BCG centre. It is fainter than the N filament, containing 10 per cent of total molecular line emission, and is disconnected from the central lump of molecular gas. The N filament is clumpy and curved in an inverted `C' shape and contains roughly 50 per cent of total molecular gas emission. Unlike S filament, it is connected to the central gas reservoir. A comparison with 5 GHz radio contours from \citet{birzan08} shows that the N filament wraps around the outer edge of the N radio jet. It is fainter at 2.4 kpc (2 arcsec) NW of the nuclear continuum source where the radio jet bends by $\sim$90$^\circ$ as shown in the middle panel of Figure~\ref {fig:hac}. The S molecular gas filament appears nearly co-spatial with the southern radio jet in projection in the sky plane. Molecular gas often forms on the edge or around radio bubbles from gas that was pushed away or lifted by the bubbles or jets \citep[see for e.g.][]{russell19}. It is possible that the molecular gas has formed around the radio jets, but it appears co-spatial in projection. The central clump of molecular gas is slightly offset from the nuclear continuum source, with its brightest emission being 0.96 kpc (0.8 arcsec) SE of the nucleus, where the radio jet bends sharply by $\sim$90$^\circ$.

There is a clear interaction between the molecular gas and the radio jets. The molecular emission peaks are nearly co-spatial with the bends in the radio jet. Either the central radio source is precessing \citep{breugel84} or the collision of the radio jets with the dense ISM at surface brightness peaks in the molecular gas changes the path of the expanding radio lobes \citep{vb85,mcnamara93,mcnamara96b}.

\subsection{Velocity structure of the molecular gas}
The velocity structure of molecular gas is fairly smooth in Abell 1795. The CO(2-1) emission in most of the regions can be well described by a single Gaussian component. However, a second component is required in two regions closer to the nucleus and in the N filament, where the radio jet takes sharp turns. The southern molecular gas filament has a smooth velocity gradient from $-$80 km s$^{-1}$ to $-200$ km s$^{-1}$ from the outer to the inner region. The FWHM is narrow in most of the filament at $\sim$40 km s$^{-1}$ but increases to $\sim$100 km s$^{-1}$ towards the inner region. The N filament has a narrower velocity gradient from $-100$ km s$^{-1}$ to $-370$ km s$^{-1}$. The FWHM is fairly constant between 60 km s$^{-1}$ to 100 km s$^{-1}$ with no clear gradient. At the end of the N filament, it increases up to 300 km s$^{-1}$. The central reservoir has velocities between $-$30 km s$^{-1}$ to 130 km s$^{-1}$ and broader FWHM between $\sim$100 km s$^{-1}$ and 250 km s$^{-1}$. The gas in the central region is dynamically distinct from the gas in the filaments. There is no indication of a rotating disk of molecular gas in the centre.

The second emission component located near the centre has velocities of $\sim$60 km s$^{-1}$ with narrow FWHM of $\sim$70 km s$^{-1}$. The two velocity components are closer in velocity but their FWHM differ by $\sim$170 km s$^{-1}$. The second emission component in the N filament is blueshifted by 50 km s$^{-1}$ compared to the first component with similarly narrow FWHM of $\sim70$ km s$^{-1}$ as the first component.

\subsection{Molecular gas mass}
\label{COmass}
    The molecular gas mass is estimated using the relation from \citet{bolatto13} as follows
    \begin{equation}
	\begin{split}
	M_{\rm mol} = & 1.05 \times 10^{4} \Bigg(\frac{X_{\rm CO}}{X_{\rm CO,Gal}}\Bigg) \bigg(\frac{1}{1+z}\bigg) \bigg(\frac{S_{\rm CO} \Delta v}{\rm Jy\, km\, s^{-1}}\bigg) \bigg(\frac{D_{\rm L}}{\rm Mpc}\bigg)^{2}\, {\rm M_{\odot}},
	\end{split}
	\end{equation}
	where $S_{\rm CO} \Delta v$ is the integrated flux density of the CO(1-0) line, $D_{\rm L}$ is the luminosity distance, $z$ is the redshift, $X_{\rm CO}$ is the CO-to-H$_2$ conversion factor and $X_{\rm CO,Gal}$ = 2$\times$10$^{20}$ cm$^{-2}$ (K km s$^{-1}$)$^{-1}$. The CO(1-0) flux density is estimated from CO(2-1) using the integrated line flux ratio CO(2-1)/CO(1-0) $\approx$ 3.2 based on similar ratios observed in BCGs \citep{russell16,vantyghem16,vantyghem17,vantyghem18}.
	
	A major source of uncertainty in the mass estimate is the uncertainty in the $X_{\rm CO}$ factor. We assume the Galactic value for this factor which has an uncertainty of about 30\% \citep{solomon87}. The true value of $X_{\rm CO}$ depends on the metallicity, temperature and density of molecular clouds \citep{bolatto13}. No direct estimates of $X_{\rm CO}$ are available for BCGs. However, the metallicities in the atmospheres of BCGs are close to the solar value and the linewidths of molecular gas clouds are similar to the molecular cloud linewidths observed in the Milky Way \citep{rose19,tremblay16}. In RXCJ0821 BCG, the molecular gas was found to be optically thick indicating abundances close to the Galactic value \citep{vantyghem17}. Therefore, we believe the Galactic $X_{\rm CO}$ factor is appropriate for this BCG as well. If the true conversion factor is similar to that of ULIRGs, the molecular gas masses would be smaller by a factor of up to 5.
	
	The total molecular gas mass in the BCG is (3.2 $\pm$ 0.2) $\times$ 10$^9$ M$_\odot$, consistent with the results of \citet{russell17b}. Nearly fifty per cent of molecular gas mass is in the N filament with a mass of (1.5 $\pm$ 0.2) $\times$ 10$^9$ M$_\odot$. The S filament has a molecular gas mass of (3.8 $\pm$ 0.4) $\times$ 10$^8$ M$_\odot$. 
	

\begin{figure*}
    \centering
    \includegraphics[width=0.8\textwidth]{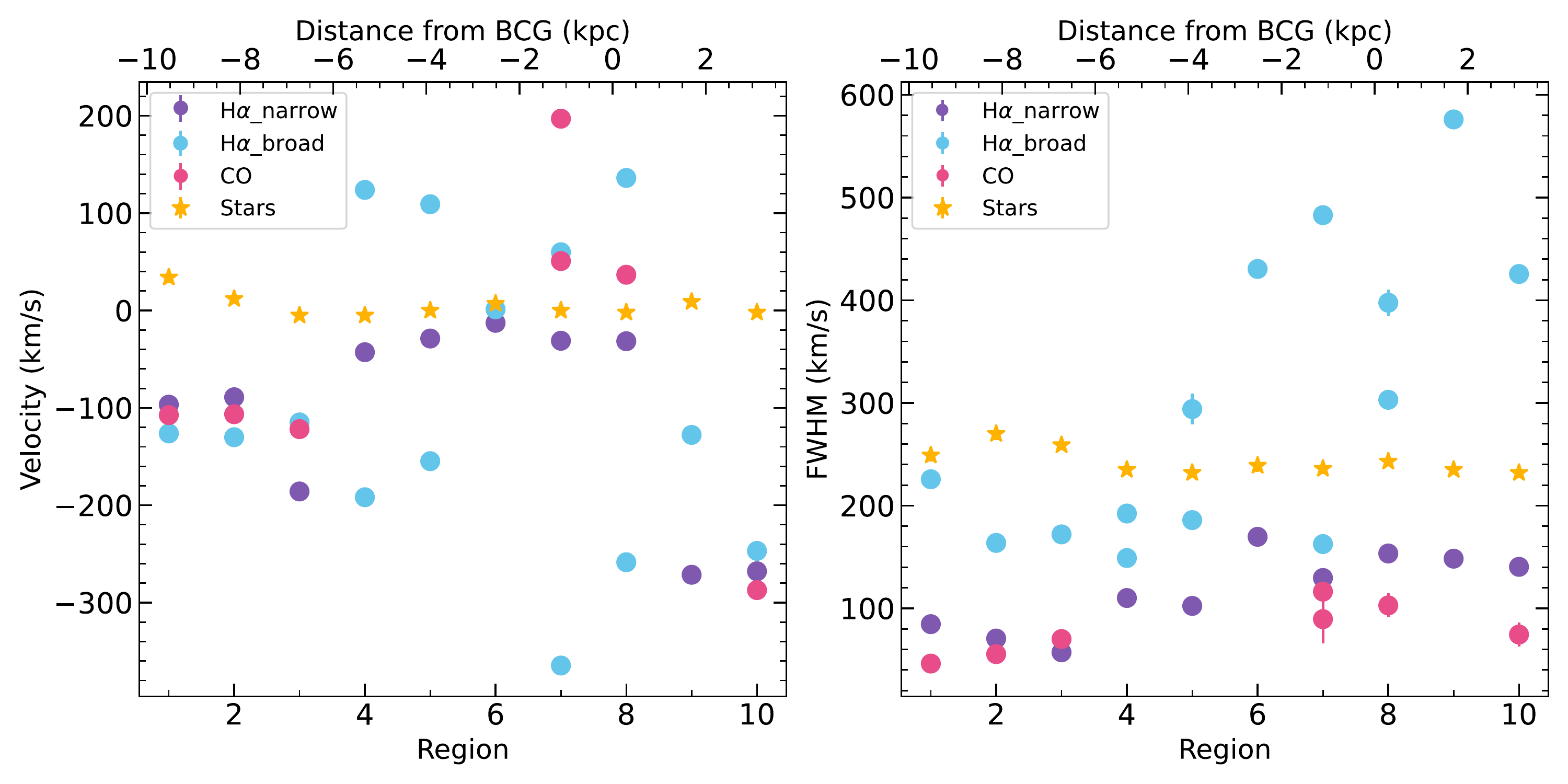}
    \caption{A comparison of the H$\alpha$, molecular gas and stellar velocities (left panel) and FWHM (right panel) in regions along the XSHOOTER slit. Stellar velocities  estimated using \textsc{ppxf} are plotted. The physical distances of regions from the nuclear radio source are shown on the top x-axis, where negative distances indicate regions to the south of the nucleus (see Fig.~\ref{fig:hac}). The molecular gas is comoving with the H$\alpha$ gas in the outer regions of filaments. The gases appear dynamically disconnected in central region.}
    \label{fig:HaCOvelComp}
\end{figure*}

\section{Comparing CO and H$\alpha$ emissions}
\label{COHacomp}

Figure 4 of \citet{olivares19} shows maps of the line of sight velocity difference, ratio of flux densities and the ratio of the velocity FWHM between molecular and H$\alpha$ gas. They showed that the flux ratio does not show strong spatial variation in the filaments indicating that the two phases are strongly coupled. In the central regions, H$\alpha$ is 2--4 times brighter compared to the filaments perhaps because of additional ionizing radiation from the AGN, young stars and interaction with the radio lobes. The ionized gas appears to be comoving with the molecular gas in most of the galaxy as indicated by a fairly constant velocity gradient. The molecular and ionized gases are very likely directly physically related where most of the ionized gas is the outer region of molecular gas clouds. Therefore, they are expected to have the same velocity structure. The FWHM of the H$\alpha$ gas is 2--4 times higher than the FWHM of the molecular gas. Similar trends are also observed in the ionized and molecular gas in other BCGs \citep{tremblay18,olivares19}. The H$\alpha$ gas is more extended than the molecular gas. Therefore, the line of sight crosses more H$\alpha$ gas than molecular gas, perhaps with different layers of gases at different speeds, resulting in a broad FWHM.

We compared the velocity and FWHM of the H$\alpha$ gas to that of the molecular gas within regions used in the XSHOOTER slit. We extracted the molecular gas spectrum and fitted it with one or more Gaussian components similar to the process described in section~\ref{Ha_kin}. In this way, we can see the effect of a line of sight on the velocities. Figure~\ref{fig:HaCOvelComp} shows the relationship between velocities and FWHM of the two gas phases along the XSHOOTER slit. The velocities of narrow components of H$\alpha$ gas are comparable to velocities of the cold molecular gas in the outer regions of both the N and the S filaments (regions 1-3 and 10). In central regions, the coupling between the two phases is more complex. The FWHM of H$\alpha$ gas is more than 10 times the FWHM of molecular gas for the broad component. The interaction between the radio jet and the ISM is strongest in central regions. Central regions also overlap with dust (see left panel of Fig.~\ref{fig:ks_new}). If the ionized and molecular gases are mixed with dust, individual layers of ionized gas along the line of sight moving at different velocities will suffer different levels of extinction depending on the location of the dust. It will affect the shape of the emission lines resulting in velocities appearing shifted compared to true their velocities. In the rest of the filament, at least one component of the H$\alpha$ gas appears coupled to the molecular gas. Thus, multiple gaussian components could be tracing the gas at different locations along the line of sight moving at different speeds. 

\section{Star formation}
\label{sf}

The bright UV emission seen in $HST$ images hints at recent or ongoing star formation in Abell 1795 \citep{mcnamara96}. The UV emission predominantly appears around radio lobes in filaments and the central region with several bright knots of UV emission. It is co-spatial with the molecular gas in the centre and in the N filament of the galaxy. However, it is offset from the molecular gas filament to the S of the galaxy. The southern molecular gas filament is broken from the central gas clump, however, the southern UV filament appears continuous (see Fig.\ref{fig:ks_new}, the middle panel). A similar trend is observed in the central reservoir where bright knots of star formation seen in the UV image is offset from the bright molecular gas emission regions.

Several studies have estimated the total star formation rate in A1795 between less than 1 $M_\odot$ yr$^{-1}$ to 20 $M_\odot$ yr$^{-1}$ \citep{mcnamara89,smith97,crawford05,hicks05,mcdonald10,donahue11} using various techniques and data at different wavelength ranges. Most star formation rate estimates using UV imaging lie between 5 and 20 $M_\odot$ yr$^{-1}$ depending on the assumed star formation history and the initial mass function (IMF). For our analysis, we estimated the UV star formation rate using an $HST$ UV images obtained from the Hubble legacy archive as described in section~\ref{hstVLAdata}.

We measured the flux of the entire UV emission region in the $HST$ F150LP image. We used Starburst99\footnote{\url{https://www.stsci.edu/science/starburst99/docs/default.htm}} code \citep{leitherer99} to generate model spectra of young stellar populations assuming solar metallicity ($Z\sim0.02$) and GENEVA 2000 tracks. We used both continuous star formation and single burst star formation histories assuming the Salpeter IMFs with slopes of 1.3, 2.3 and 3.3. We used the models with an age of 10$^7$ years, which is consistent with the sound speed rise time of inner radio bubbles of $\sim$7 Myr \citep{mcnamara96}, as well as the age of the young stellar population of 7.5$^{+2.5}_{-2.0}$ Myr in the S filament estimated using far-UV spectroscopy \citep{mcdonald14}. We applied foreground extinction to model spectra assuming a colour excess of E(B-V) of 0.0116 based on dust maps of \citet{schlafly11}. We also applied intrinsic extinction calculated from the ratio of H$\alpha$ and H$\beta$ emission line fluxes integrated over the area covering the star-forming filaments and scaled using methods described in section~\ref{extinction}. We estimated average intrinsic extinction A$_{\rm V} \sim$0.6 in the galaxy using \citet[][]{cardelli89} extinction curve. The extincted models were convolved with the $HST$ F150LP filter to establish a relationship between star formation rate and UV flux. The total SFR in 7 kpc radius aperture is 9.3$\pm$0.4 $M_\odot$ yr$^{-1}$ and 21.3$\pm$3.1 $M_\odot$ yr$^{-1}$ for continuous and instantaneous star formation histories, respectively, for IMF slope of 2.3. An IMF slope of 1.3 yields lower SFRs of 1.5$\pm$0.1 and 8.1$\pm$1.2 $M_\odot$ yr$^{-1}$, respectively, for the two star formation histories. Assuming a top-heavy IMF with a slope of 3.3 yields unrealistically high SFRs of 581 and 758 $M_\odot$ yr$^{-1}$. Therefore, the IMF slope in A1795 is most likely bottom-heavy with a slope between 1.3 and 2.3.

\subsection{Radio lobe-ISM interaction}
\label{RadioISM}

Almost all star formation knots appear on the outer regions of radio lobes providing strong evidence for star formation triggered by radio lobes \citep{mcnamara96b}. Similar spatial association between radio lobes and star formation is observed in several radio galaxies \citep[see for example,][]{duggal21}. Radio lobes can potentially compress and pressurize the gas in the interaction region and send shocks causing gravitational collapse \citep{gaibler12}. We calculated the pressure in the region of interaction between the radio lobes and the ICM following methods described in section 4.2 of \citet[][]{lacy17} (hereafter, L17). They used VLA 1.5 GHz data to estimate the pressure corresponding to the minimum energy density required to produce the observed synchrotron emission in the interaction region. We used the VLA L band image of A1795 at 1.5 GHz to calculate the mean surface brightness (0.03 Jy/beam) in the region and the radio beam size (1.2$^{\prime\prime} \times 1.1^{\prime\prime}$) while keeping all other parameters the same as in L17. We estimated the pressure in the interaction region to be (3$^{+1.5}_{-0.6}$)$\times$10$^{-10}$ dyn cm$^{-2}$, using equations 3 and 4 in L17 for path length through source of 4 kpc. The errors indicate the range of pressures for path lengths between 2 and 6 kpc. The thermal pressure of the ICM calculated from X-ray observations at 5 kpc is $\sim$4.5$\times 10^{-10}$ dyn cm$^{-2}$. It is comparable to the pressure in the interaction region suggesting that the region of interaction between radio lobes and the ICM is not overpressured and that radio lobes can only produce weak shocks. The minimum pressure required to support the molecular gas in Abell 1795 is $\sim$10$^{-8}$ dyn cm$^{-2}$ \citep{russell17b}. It is larger by at least an order of magnitude suggesting that the structure of gas is supported by magnetic fields. Thus, radio jets can create low-velocity shocks in the gas in the interaction region, that could couple more effectively to the ISM than strong shocks \citep{appleton13}.

\subsection{The role of dust}

Dust also plays a key role in cooling the gas and facilitating star formation. The left panel of Fig.~\ref{fig:ks_new} shows a strong dust lane in the galaxy along with star-forming locations. The dust is draped around the N radio lobe, co-spatial with the molecular gas in projection, perhaps entrained by the radio lobe. Typically dust and molecular gas would get destroyed due to shocks, ram pressure and the hot atmosphere of the ICM on a timescale shorter than the dynamical timescale of the outflow \citep{scannapieco15,vogelsberger19}. However, their presence suggests that they are shielded, perhaps by magnetic fields \citep[see for e.g.][]{mccourt15}. \citet{donahue11} detected Polycyclic Aromatic Hydrocarbons (PAHs) which are the smallest dust grains and molecular hydrogen lines at $\sim$8--13 $\mu$m in the rest frame using $Spitzer$ mid-IR spectrum, and also concluded that the dust is shielded from the ICM and X-ray radiation. However, dust can also grow quickly to detectable levels from cooling of warm ($<10^7$ K) gas drawn up by radio jets \citep{qiu20}. \citet{edge02} detected warm H$_2$ and FeII 1.6$\mu$m lines in their IR spectrum of the BCG. They used a larger 1.22 arcsec wide slit centred on the BCG nucleus where there is a significant amount of dust. We did not detect H$_2$ or FeII emission lines in the NIR XSHOOTER spectrum of the S filament at a level of $1 \times 10^{-16} \, \rm erg\, s^{-1}\, cm^{-2}\, \AA^{-1}$. There is no visible level of dust in the southern filament in $HST$ images. Dust could be hidden behind the stars, therefore not being visible. Nevertheless, the presence of dust in the centre and the N filament could have enhanced the cooling of molecular gas and star formation in those regions.

\subsection{Star formation efficiency}
\label{sfe}

\begin{figure*}
    \centering
    \includegraphics[width=\textwidth]{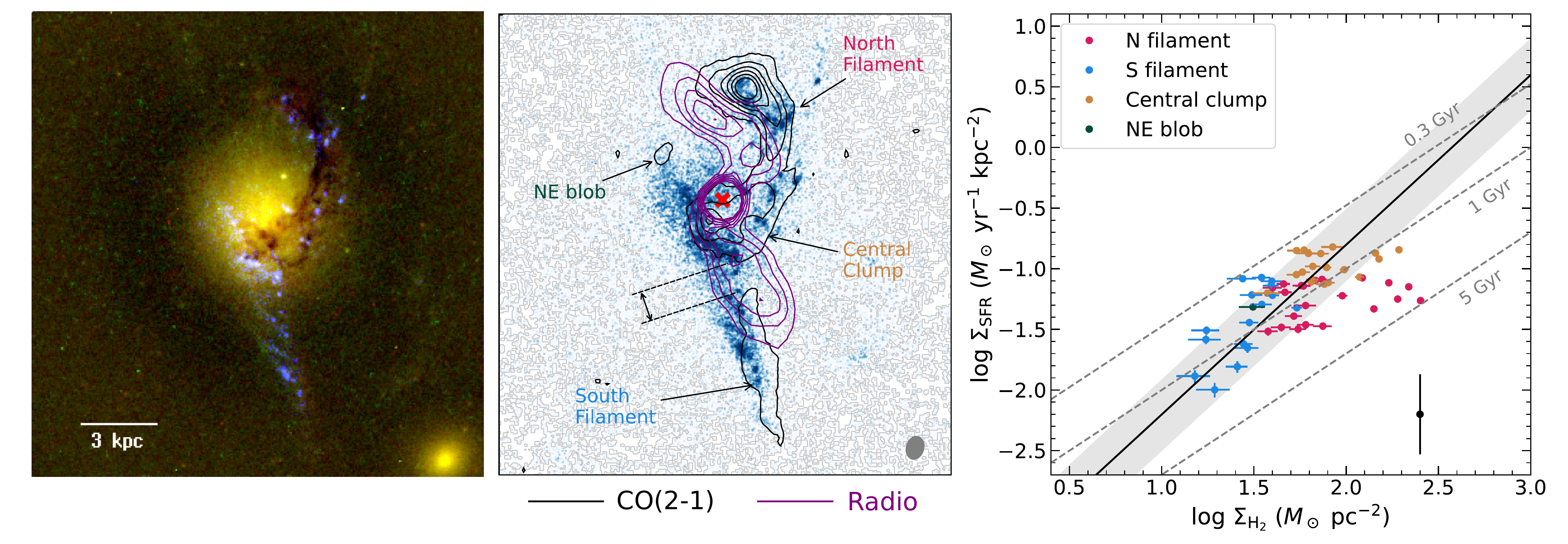}
    \caption{{\it The left} panel shows an RGB image with smoothed model subtracted $HST$F702W and $HST$F555W images in red and green, respectively, and $HST$F150LP image in blue. \textit{The middle} panel shows the $HST$F150LP image overlaid with CO(2-1) line emission contours in black and VLA X-band radio contours in purple. The offset between the molecular gas and star forming filament discussed in section~\ref{Mol_offset} is shown  between dashed black lines. \textit{The right} panel shows the relationship between $\Sigma_{\rm SFR}$ and $\Sigma_{\rm H_2}$ in ALMA beam-sized regions for different areas across the galaxy. The black line shows the global average Kennicutt-Schmidt relation and grey shaded region shows the $\pm$0.3 dex dispersion of normal galaxies from average relation. Dashed grey lines show different molecular gas depletion timescales.}
    \label{fig:ks_new}
\end{figure*}

We compared the star formation surface brightness with molecular gas surface brightness to study the efficiency of star formation. First we aligned the $HST$F555W and $HST$702W images with the $HST$150LP image such that the bright star-forming knots at the far end of the N filament visible in the UV and optical images are aligned. We then determined the offset between the central bright point source seen in the optical images which does not have a UV counterpart, with the location of the ALMA nuclear continuum source assuming they are both originating from the AGN. The offset determined also aligned the molecular gas and dust lanes very well. Then, we convolved the $HST$ image to a Gaussian kernel whose FWHM and position angle were the same as that of the synthesised beam in the ALMA image. We then re-sampled the $HST$ image onto ALMA's pixel grid using the \texttt{reproject} package in Python. The CO(2-1) and UV fluxes were calculated in ALMA beam-sized regions. The CO(2-1) flux was converted to molecular gas mass as explained in section~\ref{COmass} and SFRs were estimated from UV fluxes as explained in section~\ref{sf} assuming continuous SFH and an IMF slope of 2.3.

The right panel of Fig.~\ref{fig:ks_new} shows the relationship between star formation rate surface density $\Sigma_{\rm SFR}$ and molecular gas surface density $\Sigma_{\rm H_2}$ in different parts of the galaxy estimated by dividing the SFR and molecular gas mass by beam area in appropriate units, respectively. Most of the regions follow the the Kennicutt-Schmidt (KS) relation \citep{kennicutt98} shown as a black line. The KS relation is a well characterized relationship between $\Sigma_{\rm SFR}$ and $\Sigma_{\rm H_2}$ in nearby galaxies given by $\Sigma_{\rm SFR} \propto \Sigma_{\rm HI + H_2}^{1.4}$. We estimated average molecular gas depletion times due to star formation ($\tau_{\rm depl,SFR} = {\rm M}_{\rm H_2}/{\rm SFR}$) of 1 Gyr assuming star formation continues at 9.3 $M_{\odot}$ yr$^{-1}$. The $\tau_{\rm depl, SFR}$ in the S filament and the central clump is $\sim$ 0.8 and 0.9 Gyr, respectively. It is twice as long in the N filament at 1.7 Gyr. These results indicate a lower star formation efficiency than circumnuclear starbursts in nearby galaxies with similar gas densities.

A few regions in the N filament lie below the KS relation. They are located directly along the radio jet path where the radio jet bends around the brightest molecular gas emission in the N filament. In these regions the depletion time approaches 5 Gyr indicating a much lower star formation efficiency. It is consistent with low efficiency of jet-induced star formation observed in other galaxies \citep[see for e.g.,][]{salome16,zovaro19}. Alternatively, either the molecular gas mass is overestimated or the star formation is underestimated. The molecular gas mass can be overestimated if the value of the $X_{\rm CO}$ conversion factor is lower than its Galactic value. The $X_{\rm CO}$ factor depends primarily on the density, metallicity and temperature of the molecular gas. In highly starforming regions, the $X_{\rm CO}$ is up to a factor of five below its Galactic value, due to high gas temperature and velocity dispersion \citep{bolatto13}. These regions have high molecular gas FWHM suggesting that molecular gas masses could have been overestimated. On the other hand, star formation can be underestimated if the underlying IMF is top-heavy. Further detailed studies are required to draw a firm conclusion.

The star formation depletion timescales are much longer than the age of the stellar population and the radio lobes. However, most of the molecular gas in Abell 1795 is flowing at a rate of $\sim$ 54 M$_\odot$ yr$^{-1}$ \citep{tamhane22}. If the gas is inflowing, it can fuel a starburst in the center, although it is unlikely that the molecular gas filaments are gas inflows \citep{russell17b}. If the gas is being driven out of the galaxy, the molecular gas in the galaxy will be depleted within $\sim$6$\times$10$^7$ years, which is much shorter than $\tau_{\rm depl, SFR}$. Thus, although radio jet-ISM interaction may have triggered star formation, this form of star formation has lower efficiency and may be quenched eventually by the negative feedback of radio jets.

\subsection{The offset between star formation and molecular gas}
\label{Mol_offset}

The region of disconnection between the S molecular gas filament and the central clump mentioned in section~\ref{molmorph} and shown in the middle panel of Fig.~\ref{fig:ks_new} is coincident with bright H$\alpha$ and UV emission. It shows an offset between star formation and molecular gas. Overall, the southern molecular gas filament is offset from the star-forming filament by $\sim$0.4 kpc in the perpendicular direction if the molecular gas filament were to extend to the centre and $\sim$1 kpc in the direction of the filament. Similar offsets between star formation and cold gas have also been found in NGC1275 and simulations of AGN feedback \citep{canning14,li15}. We discuss two possibilities when this can happen.

First, the ram pressure of the expanding radio lobes has most likely triggered star formation in Abell 1795. The star formation may have consumed the molecular gas in the inner region of the S filament and its NE extension. Supernova feedback from young stars can also destroy the molecular gas around them. Assuming that the entire region of the star-forming filament had a similar molecular gas surface density as that observed in the S molecular gas filament when star formation began, the required star formation rate to consume the molecular gas in the inner parts of the S filament in $\sim$10$^7$ years is 70 M$_\odot$ yr$^{-1}$. It is 5--7 times larger than observed star formation rates. At the observed SFR, it would take $\sim 10^{9}$ yrs to consume the molecular gas in that region, which is about two orders of magnitude longer than the age of radio lobes and an order of magnitude longer than the ages of inner X-ray cavities \citep{kokotanekov18}. Even if we assume that the star formation proceeded at a higher rate than observed, it is not clear why gas in other parts of the S filament has not yet been consumed. If we assume that the ram pressure due to radio lobes is not strong enough in the outer part of the molecular gas filament to trigger star formation, it still does not explain the observed level of molecular gas in the N filament. 

Alternatively, the stars formed in the molecular gas filament have decoupled from the flow and are moving in the gravitational potential of the galaxy without significant resistance, whereas the molecular gas is acted upon by the ram pressure from the radio lobes and the ICM. It is similar to gas stripping in jellyfish galaxies, where the ram pressure of the ICM strips the gas from infalling galaxies while the stars in the galaxy are unaffected \citep[see for example,][]{fumagalli14,sun22}. If the S filament is an inflow, stars would always lead the gas, as the gas would slow down due to the ram pressure of the ICM and radio jet. It can also explain the extension of the UV emission to the NE of the BCG nucleus, where the stars falling ballistically would overshoot the nucleus and are moving in that region. In this scenario, the stars are leading the gas by $\sim$1 kpc. For a $\sim$7 Myr old stellar population to have 1 kpc separation from the molecular gas, it would need to have an initial velocity of $\sim$400 km s$^{-1}$ in the rest frame of the BCG, following arguments presented in section 4 of \citet{li18}. It is higher than the stellar velocity dispersion in the BCG of 297 km s$^{-1}$ but close to the average velocity of cluster galaxies of 368 km s$^{-1}$ relative to the BCG. A high-resolution far-UV spectrum of the inner filament where the contribution of old stars and nebular emission lines would be minimal is required to obtain the velocities of young stars in the filament.

\section{Emission line ratios}
\label{elineratios}
As discussed in section~\ref{ionizedgas}, the products of \textsc{PyParadise} allowed us to make maps of emission line fluxes. We use these maps to determine the excitation state of the ionized gas, its density and dust extinction in the galaxy as discussed in the sections below. We smoothed the MUSE data cube by a Gaussian kernel with FWHM of 1.3$^{\prime\prime}$, equal to the seeing during the observations to reduce the pixel-to-pixel noise and fit spectra extracted from each spaxel using \textsc{PyParadise} as discussed in section~\ref{ionizedgas}. In the following analyses, pixels with S/N $>3$ for respective nebular emission lines were used.

\subsection{Extinction map}
\label{extinction}

\begin{figure}
    \centering
    \includegraphics[width=0.45\textwidth]{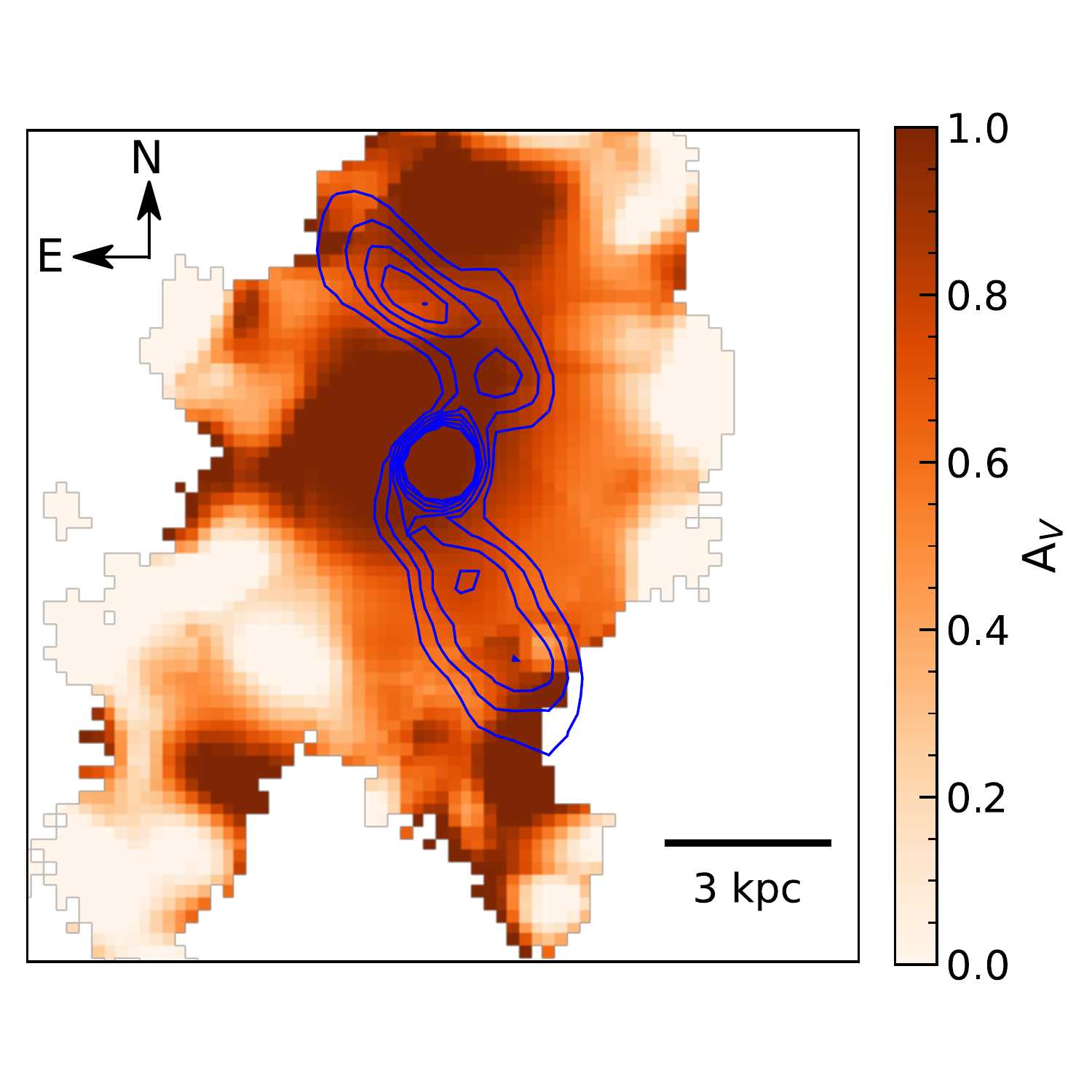}
    \caption{A map of extinction in the visual band ($A_V$) calculated as explained in section~\ref{extinction}. Radio contours in the VLA X-band are overlaid in blue.}
    \label{fig:extinction}
\end{figure}

Abell 1795 hosts a very prominent dust filament in the N roughly co-spatial with the molecular gas as shown in the left panel of Fig.~\ref{fig:ks_new}. Therefore we expect significant dust extinction in the galaxy. We estimated the amount of visual extinction ($A_V$) by making a map of the ratio of H$\alpha$/H$\beta$ Balmer decrement and scaled it to calculate E(B$-$V) as follows \citep{dominguez13}:
\begin{equation}
    E(B-V) = \frac{2.5}{k(\lambda_{\rm H\beta}) - k(\lambda_{\rm H\alpha})} {\rm log_{10}} \left[ \frac{\rm (H\alpha/H\beta)_{obs}}{\rm (H\alpha/H\beta)_{int}} \right],
\end{equation}
where $k(\lambda_{\rm H\beta}) \approx 3.7$ and $k(\lambda_{\rm H\alpha}) \approx 2.63$ are extinction curves evaluated at H$\beta$ and H$\alpha$ wavelengths, respectively, given by \citet{cardelli89} and we assume (H$\alpha$/H$\beta$)$_{\rm int} \approx 2.86$ for an electron temperature of T$=$10$^4$ K and electron density of $n_e = 10^2$ cm$^{-3}$ for Case B recombination. $A_V$ is then simply calculated as $R_V \times$ E(B$-$V), where $R_V = 3.1$.

In Figure~\ref{fig:extinction}, we show the map of $A_V$ in the BCG. As expected, there is a significant amount of extinction in the centre and the N filament with $A_V$ exceeding 1 mag. The S filament has much lower extinction, indicating a low amount of dust compared to the N filament. It is consistent with the absence of a visible level of dust in the optical image of the S filament (see left panel of Fig.~\ref{fig:ks_new}), as well as the non-detection of dust emission features in the NIR XSHOOTER spectrum of the S filament. Again we note that if the warm ionized gas is on the near side of dust along the line of sight, it will not suffer from as much extinction as it would if dust and ionized gas are mixed or if the ionized gas is on the far side of dust in the S filament. The average extinction calculated by taking the mean of pixels in the region covering both filaments is 0.6. Interestingly, the dust in the centre is predominantly to the NE of the nucleus, roughly co-incident with the region that could be experiencing a shock, assuming the extinction map traces dust in the galaxy. Also, the north radio lobe appears to bend around the north blob of dust. It is also the location of the brightest CO emission indicating a large amount of molecular gas. Therefore it appears that the dust and gas have changed the direction of the radio lobes. We use this extinction map to deredden emission line fluxes in the galaxy in a similar way as in the process described in section~\ref{sf}. In the following sections, emission line fluxes in each spaxel are corrected by the level of extinction measured in that spaxel. It is not clear why the region of large $A_V$is offset from the centre and from the dust lane to the S and W of the central radio source seen in the left panel of Fig.~\ref{fig:ks_new}, but coincident with the arc of high FWHM.

\subsection{Electron density}
\label{edensity}

\begin{figure}
    \centering
    \includegraphics[width=0.45\textwidth]{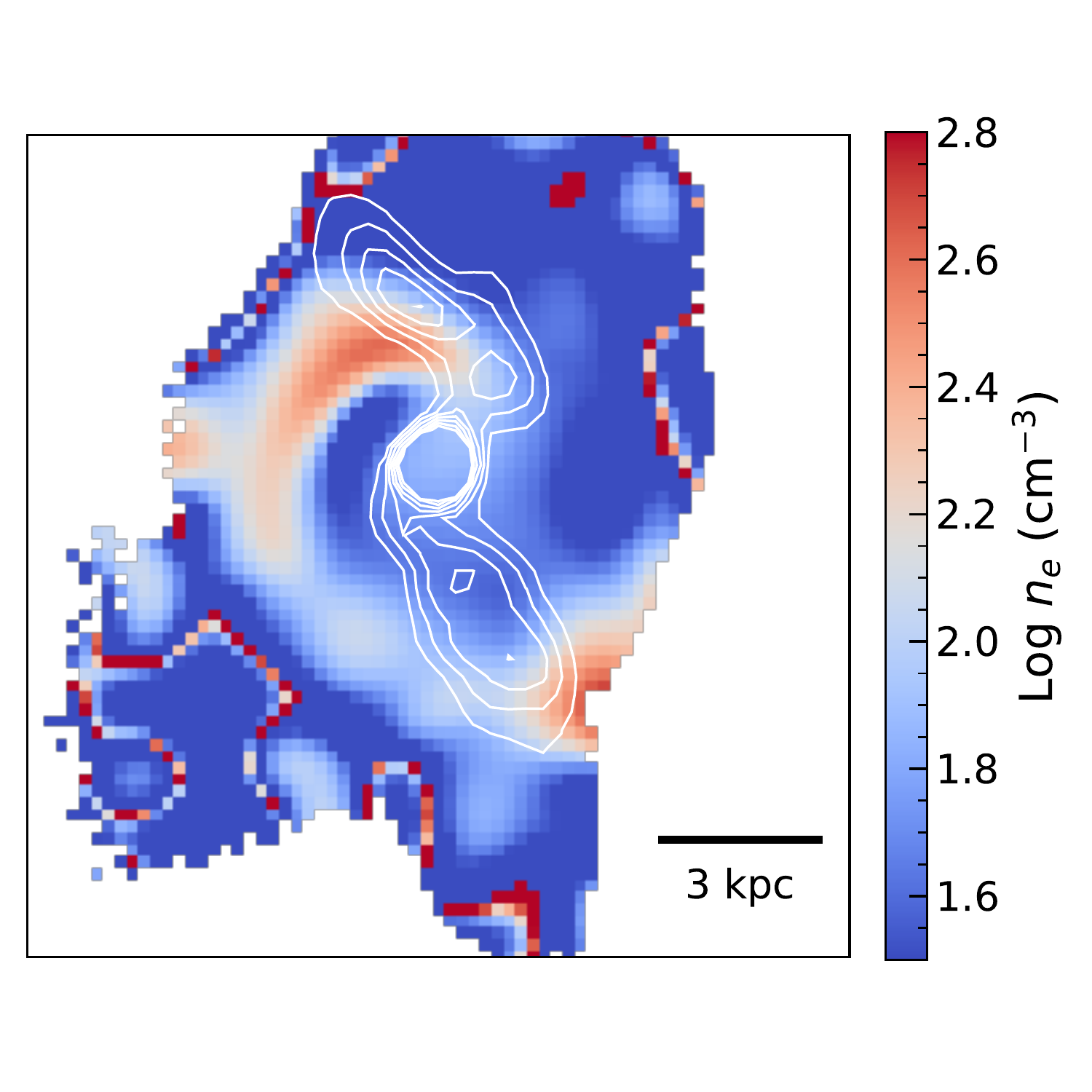}
    \caption{The figures show an electron density map derived from [SII]$\lambda$ 6717 / [SII]$\lambda$ 6732 as explained in section~\ref{edensity}. Radio contours in the VLA X-band are overlaid in white. Pixels where densities are below 100 cm$^{-2}$ are upper limits. The artifacts around the edges are artefacts due to smoothing the original data cube.}
    \label{fig:edensity}
\end{figure}

The high-resolution spectra allow us to resolve the [SII] doublets. The relative intensities of [SII]$\lambda$ 6716 to [SII]$\lambda$ 6731 can be used to estimate electron density ($n_e$) in the gas, since the doublets are sensitive to the effects of collisional de-excitation and insensitive to variations in electron temperature, due to their similar excitation energies. Therefore, their excitation rate depends on the ratio of the collision strengths and the ratio of line intensities depends only on the density of the gas \citep{osterbrock89}. We scaled the [SII]$\lambda$ 6716/[SII]$\lambda$ 6731 ratios to get estimate $n_e$ following the procedure given in \citet{proxauf14}, where we assumed electron temperature $T_e \sim 10^4$ K. This method is only sensitive for gas densities between 100 and 10$^{4}$ cm$^{-3}$. In Figure~\ref{fig:edensity}, we show the $n_e$ map of the BCG. The electron densities in the filament are $\sim 100$ cm$^{-3}$ and potentially lower, as for densities lower than 100 cm$^{-3}$, [SII] line ratio becomes ineffective and can only provide an upper limit. In the center, $n_e \sim 100$ cm$^{-3}$. It is consistent with previous studies and [SII] line ratios found in other BCGs \citep{crawford99,hamer16}. In the N filament and outer regions of the galaxy, gas density is potentially lower than 100 cm$^{-3}$ indicated by dark blue regions. However, we note that the [SII]$\lambda6731$ line is close to the atmospheric absorption band and may have been affected by atmospheric absorption resulting in higher line ratio and lower gas densities. To the N-NE of the centre, on the outer edge of the central radio source, the gas has higher electron densities of $\sim$400 cm$^{-3}$. This feature is co-spatial with the high FWHM arc shown in Figure~\ref{fig:HaMaps}. As discussed in section~\ref{Ha_kin}, it could either be a result of beam smearing on two separate gas filaments moving at different speeds superimposed on the sky in projection along the line of sight or it indicates a shock front. If it is a shock, the ratio of densities between the post and pre-shocked gas is greater than 1. Weak shocks with a post to pre-shock density ratio of up to 2 have been observed in the ICM of galaxy clusters \citep{sanders06,fabian06}. If we assume similar density ratio of 2 in the ISM and the ratio of specific heats $\gamma$ of 5/3 for the gas yields a Mach number of 1.7 using equation (10-20) of \citet{spitzer78}, indicating a weak shock. Pressures in the region of interaction between the radio jets and the ISM also indicate weak shocks as discussed in section~\ref{RadioISM}. Therefore, the arc of elevated densities and FWHM likely indicates a shock front driven either by radio jets or the peculiar motion of the BCG. However, alternative scenario of superposition of two filaments along the line of sight cannot be completely ruled out. Higher spatial resolution observations are required to minimize the effect of beam smearing and for a better understanding of the feature.

\subsection{Excitation state}
\begin{figure}
    \centering
    \includegraphics[width=0.45\textwidth]{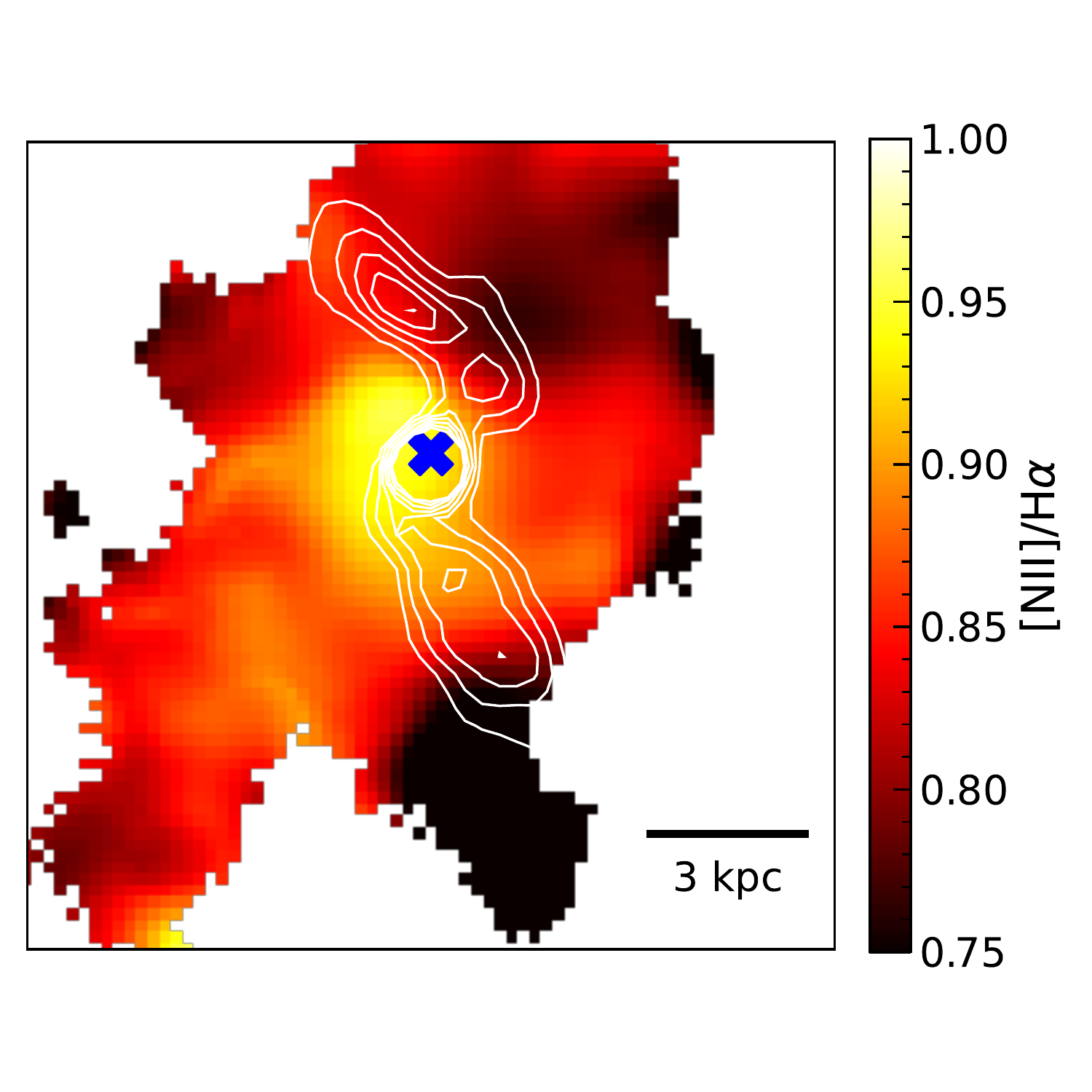}
    \caption{The ratio of [NII]/H$\alpha$ represents the excitation state of the ionized gas. The blue cross denotes the location of the nuclear radio source. The VLA X-band radio contours are overlaid in white.}
    \label{fig:exState}
\end{figure}

The [NII] to H$\alpha$ ratio can probe the excitation state of the gas due to its sensitivity to AGN. Generally, star formation has a softer radiation field than AGN. Thus, changes in the [NII]/H$\alpha$ ratio in a galaxy indicate different gas excitation mechanisms in different parts of a galaxy. In Figure~\ref{fig:exState}, we show the [NII]/H$\alpha$ map of the BCG. In many parts of the BCG surrounding the nucleus, the ratio is $\sim$0.85. At $\sim$4 kpc south of the nucleus, the ratio changes sharply from $\sim$0.85 to less than 0.7. The transition occurs at the tail end of the `V'-shaped redshifted wings observed in H$\alpha$ velocity map shown in the left panel of Fig.~\ref{fig:HaMaps}. The reason for sharp change in the excitation ratio is not entirely clear. The ratio is highest in the centre of the BCG and in the region of possible shock. The trend of high [NII]/H$\alpha$ ratio in the centre compared to outskirts is also observed in other BCGs \citep{hamer16}. It is expected as the gas in the central region is excited by more energetic sources perhaps by shocks or AGN photoionization.

\begin{figure*}
    \centering
    \includegraphics[height=0.9\textheight]{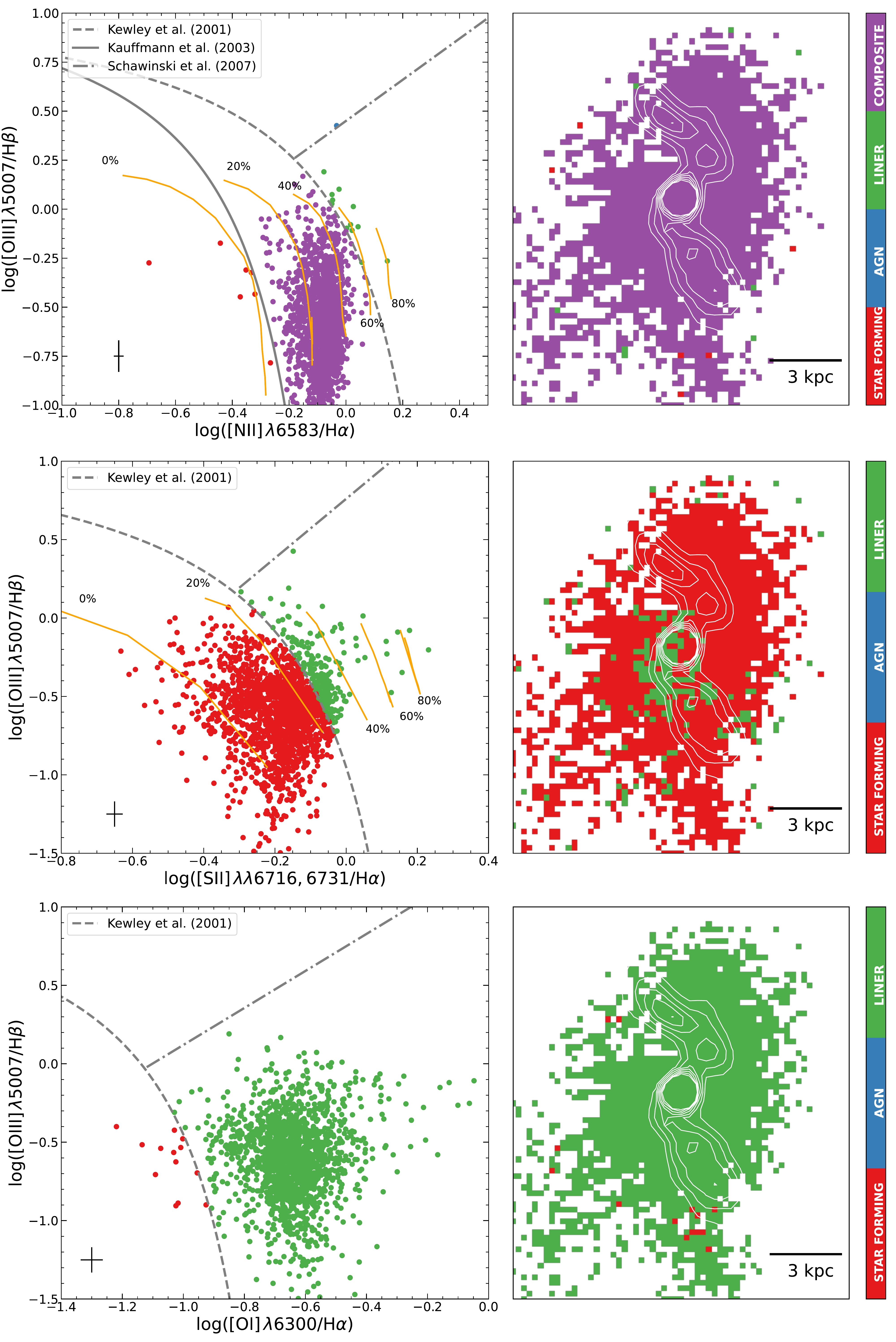}
    \caption{The figure displays the emission line diagnostic diagrams. The left panel of the top, middle and the bottom row shows the standard BPT diagnostic diagrams using the [OIII]$\lambda$5007/H$\beta$ plotted against [NII]$\lambda$6583/H$\alpha$, [SII]$\lambda\lambda$6716,6731/H$\alpha$ and [OI]$\lambda$6300/H$\alpha$ line ratios, respectively. The right panel in each row displays the location of spaxels on the galaxy colour-coded by the well-known theoretical classification boundaries in BPT diagrams \citep{kewley01,kauffmann03,schawinski07} shown as grey dashed and dotted lines in the left panels. We also show in the top left panel composite star formation + shock ionization models of \citet{mcdonald12} as orange curves with the contribution of slow shocks indicated next to the curves. Radio contours at 8 GHz are shown in white. Spaxels with S/N$>$3 for all lines were included.}
    \label{fig:museBPT}
\end{figure*}

\subsection{Optical diagnostic diagrams}

The broad wavelength coverage of the spectrum allows us to use ratios of various gas emission lines to try to determine gas excitation mechanisms. The most commonly used diagnostics line ratios are [NII] $\lambda$6583/H$\alpha$, [OI] $\lambda$6300/H$\alpha$, [SII] $\lambda$6716, 6731/H$\alpha$, and [OIII] $\lambda$5007/H$\beta$, the so-called BPT diagnostic diagrams \citep{baldwin81}. They are generally used to approximately differentiate between different excitation mechanisms. The strength of each nebular emission line in each spaxel of the data cube was measured as described in section~\ref{elineratios} and S/N cut was applied. The ratio of line fluxes was calculated in each pixel and was used to produce the BPT diagram of the BCG. Figure~\ref{fig:museBPT} shows BPT diagnostic diagrams of the BCG. Regions predominantly populated by star-forming, AGN and composite galaxies from the Sloan Digital Sky Survey (SDSS) as identified by \citet{kauffmann03,kewley01} and \citet{schawinski07} are shown by grey solid, dashed and dash-dotted lines, respectively. The points are coloured by the regions in which they sit. The right panels show the distribution of points in the BPT diagram on the sky. We note that the [SII]$\lambda$6731 flux could have been underestimated leading to more regions represented as having star formation as a primary gas ionization mechanism. Additionally, we ignored spatial variation in extinction in emission line fluxes introducing additional scatter.

The vast majority of points have a composite or low-ionization nuclear emission-line region (LINER)-like ionization sources involving [NII] and [OI] lines as shown in the left panel of the first and the third row. The composite and LINER-like points are uniformly distributed throughout the BCG with no variation, similar to the findings in Abell 2597 \citep[see][Fig. 15]{tremblay18}. However, the BPT diagram involving the [SII] (second row in Fig.~\ref{fig:museBPT}) line shows that the majority of points have star formation as their primary source of ionization. The [NII]/H$\alpha$ ratio also indicates extreme starburst-like ionization according to \citet{kewley01} line. These results are consistent with the analysis presented in \citet{mcdonald12}, where [OIII]/H$\beta$ ratios are lower and/or [NII]/H$\alpha$, [SII]/H$\alpha$, and [OI]/H$\alpha$ ratios are higher compared to SDSS galaxies used to define different ionization regions. Interestingly, the region positioned along the outer edge of the sharp bend in the S radio jet has LINER-like ionization with [SII]/H$\alpha$ line as shown by green points in the middle right panel. Perhaps the gas in that region is predominantly ionized by shocks driven by the radio jet. Clearly, the classification in BPT diagrams for different line ratios is not unambiguous. However, the presence of strong UV emission in the BCG clearly indicates recent star formation. The situation is more complex than ionization by star formation.

We also plotted the ``slow shock + stellar photoionization" models of \citet{mcdonald12} as orange lines with a varying contribution of slow shocks in gas ionization. In the middle panel involving the [SII] line, less than 20 per cent contribution comes from shocks in most of the galaxy except in the central region where it is $\sim30$ per cent. The [NII]/H$\alpha$ ratio is more sensitive to shock ionization \citep{kewley02}. The top panel involving the [NII] line shows that the majority of points have 20-40 per cent contribution from shocks, consistent with results of \citet{mcdonald12}, where they find less than 40 per cent contribution from shocks in gas ionization. Other mechanisms such as conduction \citep{sparks12}, cosmic ray heating \citep{ferland09,fabian11,johnstone12} and thermal radiation from the cooling flow \citep{polles21} may also contribute at the same time with varying relative contributions depending on the location in the galaxy.

\section{Velocity structure function in Abell 1795}
\label{vsf}

It is clear that the radio lobes are interacting and influencing the motion of cold and ionized gases. The radio lobes can drive turbulence in the ISM of the BCG which can produce local density fluctuations that may seed star formation \citep{krumholz05}, while simultaneously supporting the gas against gravitational collapse at larger scales \citep{klessen00}. One common way to study turbulence is through the velocity structure function (VSF) which is a statistical tool to access the properties of velocity fluctuations. We computed the first order VSF of the H$\alpha$ and molecular gas by first masking pixels with velocity errors greater than 10 km s$^{-1}$. We only considered 16$^{\prime\prime} \times 16^{\prime\prime}$ region centred on the BCG that covers all molecular gas filaments and inner H$\alpha$ gas filaments. The velocity maps in Fig.~\ref{fig:vsfmaps} show the regions used to calculate the VSF for H$\alpha$ and molecular gas. For each velocity map, we compute the VSF by calculating the absolute difference between the velocities of each unique pair of pixels as $|\delta v|$ and the projected separation between those pixels ($\mathscr{l}$). We then sorted pixel pairs by $\mathscr{l}$, binned the data with 10$^4$ pixel pairs per bin and averaged the data in groups of $\mathscr{l}$. There is a small difference between the spatial resolutions of the ALMA and MUSE data. For MUSE data spatial resolution refers to the seeing of 1.36$^{\prime\prime}$. We smoothed the ALMA image to the seeing of MUSE observations and computed the VSF to examine the effect of the resolution. We found that for low-resolution VSF, power at all scales is suppressed marginally ($<0.1$ dex) meaning that the VSF curve lies systematically below the VSF for native resolution, while shape of the VSF does not change significantly. We use the VSF with the native resolution for both ALMA and MUSE data.

\begin{figure*}
    \centering
    \includegraphics[width=1.1\textwidth]{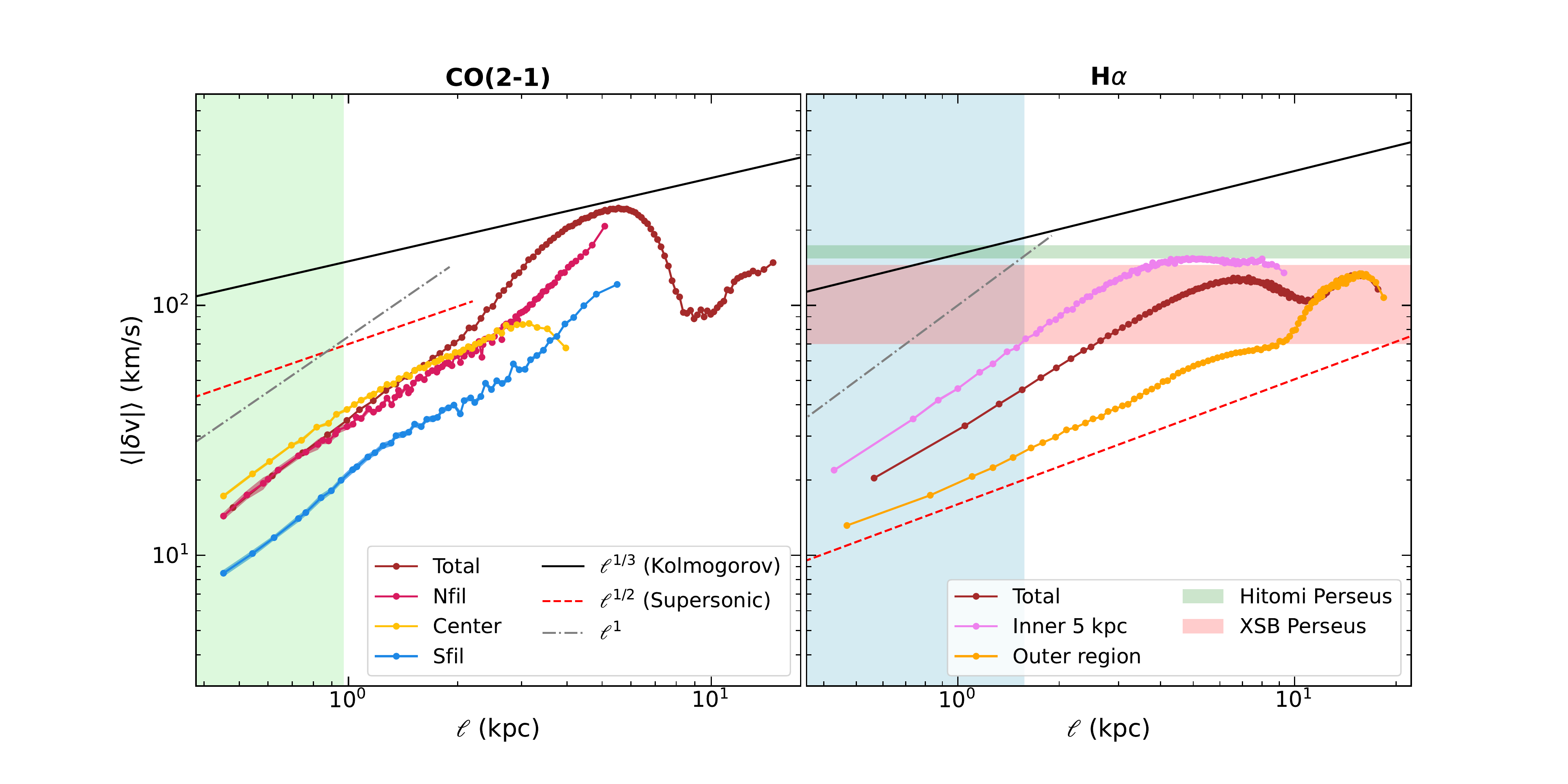}
    \caption{The figure shows the velocity structure function of the cold molecular gas (left panel) and warm ionized gas (right panel) traced by ALMA CO(2-1) and MUSE H$\alpha$ observations, respectively. The thickness of the lines indicates the errors in each bin. To guide the eye, we show the solid black line with a slope of 1/3 for Kolmogorov turbulence, a dashed red line with a steeper slope of 1/2 for supersonic turbulence and a grey dash-dotted line representing a linear relationship between the velocity scale and the spatial scale $\mathscr{l}$. Spatial scales smaller than the spatial resolution of MUSE and ALMA are shown by shaded blue and green vertical regions, respectively. Different regions used in calculating the VSF are shown in Fig.~\ref{fig:vsfmaps}.}
    \label{fig:vsf}
\end{figure*}

In the left panel of Figure~\ref{fig:vsf}, we show the VSF for the molecular gas. The total VSF for the molecular gas peaks and flattens at a pair separation of $\sim$5 kpc, corresponding to $|\delta v| \sim$250 km s$^{-1}$. The peak is most likely caused by the velocity difference between the central clump and the outer edge of the N molecular gas filament, separated by a projected distance of $\sim$5.5 kpc. The VSF of the total H$\alpha$ gas in the BCG flattens at $\sim$ 7 kpc scale corresponding to $|\delta v| \sim$120 km s$^{-1}$, as shown in the right panel of Fig.~\ref{fig:vsf}. The length scale corresponding to the VSF peak may indicate the driving scale of turbulence, which is comparable to the size of the radio lobes. For reference, we show the turbulent velocity of the X-ray gas inferred from X-ray surface brightness fluctuations \citep{zhuravleva14}, and turbulent velocity measured by Hitomi for the Perseus cluster \citep{hitomi16}, respectively, indicated by red and green shaded regions in the right panel. The similarity of turbulent velocities of hot and cold phases suggests that radio-mechanical feedback can drive turbulence in cold filaments.

The H$\alpha$ VSF has a slope of $\sim$1/2 until it peaks. At scales between 1 and 3 kpc, the molecular gas has a slope of $\gtrsim$1/2, which is characteristic of supersonic turbulence \citep{federrath13} mixed with a smooth velocity gradient \citep{hu22}. Above 3 kpc scales, the VSF of individual molecular gas filaments becomes steeper with a slope of 1, indicating that the velocity gradient dominates at these scales. However, for the total molecular gas VSF, the steepening starts at $\sim$2 kpc with a slope exceeding 1. This is likely due to the N filament intersecting the central gas clump. An extreme example of such steepening would be observed in an ordered motion similar to that in a circumnuclear disk with a radius of a few kpc and velocity shifts of $\sim\pm$100 km s$^{-1}$. It can dramatically increase the VSF amplitude at small separations.

The VSFs of the filaments have been examined in other BCGs such as Perseus, Abell 2597 and Virgo galaxy cluster \citep{li20,hillel20}. The observed VSF slope of 1/2, characteristic of supersonic turbulence, indicates the turbulent motion of cold gas driven by X-ray cavities or radio lobes. However, analysis of individual filaments in the Perseus cluster, accompanied by numerical simulations \citep{qiu21,hu22} showed that velocity gradients characterized by a VSF slope of 1 may dominate at scales larger than a few kiloparsecs, especially viewed from an angle close to the flow direction. In the case of Abell 1795, therefore, the VSF is likely dominated by the laminar motion of the filaments and is limited by the observed size of the filaments. The large observed filaments are likely made of smaller unresolved filaments or sheets of molecular gas. Higher-resolution observations probing smaller scales are required to probe the nature of turbulence in the molecular phase.

Although the power-law slope of the VSF is often attributed to turbulence, some studies have suggested that it could also be a result of laminar gas flows and large eddies formed in the gas being uplifted behind bubbles blown by radio jets \citep{zhang22}. Therefore, caution needs to be taken when interpreting the VSFs.
In addition to laminar flows and orientation effects, the presence of magnetic fields can suppress turbulence and steepen the slope of the VSF \citep{bambic18,mohapatra22}. Multiple indicators suggest that the cold gas filaments are magnetized as discussed in the paper. Therefore, magnetic fields may also contribute to steepening of the VSF in Abell 1795. Nevertheless, the H$\alpha$ gas VSF slopes, flattening scale and turbulent velocities point to supersonic turbulence mixed with velocity gradients and orientation effects. Large statistical studies of VSF of many systems sampled by different lines of sight may be able to provide a better understanding of the VSF. If we assume that the VSF is characterised by supersonic turbulence and the turbulence energy cascades down from large scales to the scales of giant molecular clouds (GMCs) triggering star formation, the resulting star formation efficiency is expected to follow the KS-relation \citep{krumholz05}. The star formation in A1795 follows the KS relation as shown in~\ref{sfe}. However, simulations suggest that the star formation can also generate supersonic turbulence in GMCs on sub-pc scales and suppress its efficiency over time \citep{hu22a}. We are unable to probe the turbulence at the scales of GMCs. We tentatively find that on a timescale of a few million years, radio jets can drive supersonic turbulence which can trigger star formation locally with a low-efficiency.

\section{Conclusions} \label{conclusion}

We studied the interaction between radio jets and the ISM and the nature of radio-jet triggered star formation in Abell 1795 by combing  VLT XSHOOTER, MUSE, ALMA and $HST$ data. We determined the star formation rate using the UV $HST$ images of the BCG and star formation efficiency by combining it with the ALMA molecular gas data. Using the nebular gas emission line ratios derived from MUSE IFU data, we were able to determine the level of extinction, electron density, excitation level and ionization mechanism in the BCG. We also detected a shock-like feature $\sim$2 kpc N-NE of the nucleus. The results are summarised below.

\bigskip
 \noindent
 $\bullet$
 We measured extinction corrected ongoing UV star formation rate of 9.3 M$_\odot$ yr$^{-1}$ in the BCG. The star formation rate has a low efficiency with an average depletion time of $\sim$1 Gyr. However, the efficiency of star formation is location dependant and it is even lower in the region of interaction between radio jets and the gas. The offset between star formation and the molecular gas in the S filament suggests that the stars have decoupled from the molecular gas and are moving in the gravitational potential of the BCG while the gas motion is being continually affected by the radio jets and the surrounding medium.

 \bigskip
 \noindent
 $\bullet$ 
 The ionized gas traced by the H$\alpha$ emission line is more extended than the molecular gas. The BCG has a total H$\alpha$ luminosity of (4.180$\pm 0.002$)$\times 10^{41}$ erg s$^{-1}$, corresponding to ionized gas mass of (1.640$\pm$0.001)$\times10^7$ M$_\odot$. The velocity map of H$\alpha$ reveals a `V'-shaped wing of redshifted gas wrapped around the southern radio lobe with velocities of $\sim$100 km s$^{-1}$. In other parts of the BCG, the ionized gas is blueshifted to velocities lying between $-100$ to $-370$ km s$^{-1}$, similar to the velocities of molecular gas, suggesting that the two gas phases are co-moving. The FWHM map reveals an arc of high FWHM with values reaching $\sim$700 km s$^{-1}$ to the N-NE of the nucleus. Other areas of the BCG have FWHM of $\sim$250 km s$^{-1}$, roughly 2-3 times the FWHM of the molecular gas.
 
 \bigskip
 \noindent
 $\bullet$
 We measured an average extinction of 0.6 in the star-forming filaments in the BCG using H Balmer decrement. The electron density in the ISM traced by [SII] line ratios in the BCG is $\sim100$ cm$^{-3}$ or lower. Electron densities in the region cospatial with the high FWHM arc are 4 times higher compared to surrounding regions which could indicate a weak shock perhaps driven by radio jets or the peculiar motion of the BCG. The standard nebular emission line diagnostic diagrams classifications are ambiguous, however, they are represented by models of gas ionization with star formation and a small contribution from slow shocks.
 
 \bigskip
 \noindent
 $\bullet$
 We also computed the VSF of the molecular and ionized gases. The molecular gas VSF is dominated by the size of the filaments and laminar motion with slopes of $\sim$1. The H$\alpha$ VSF slopes of $\gtrsim$1/2, turbulent velocities of $\sim$120 km s$^{-1}$ and flattening scale of $\sim$7 kpc similar to the size of radio lobes tentatively indicate that radio-mechanical feedback can drive supersonic turbulence in the ionized gas.
 
 \bigskip
 \noindent
 $\bullet$
 The XSHOOTER spectrum of the south star-forming filament is dominated by strong nebular emission lines and older stars in the galaxy making it unfeasible to determine the velocity of young stars. We did not detect NIR H$_2$ emission lines tracing warm molecular gas at a level of $\sim$1$\times 10^{-16}$ erg s$^{-1}$ cm$^{-2}$ \AA$^{-1}$ in the NIR XSHOOTER spectrum of the S filament.\\

We presented a detailed analysis of extended radio-jet triggered star formation. Abell 1795 presents an excellent example of both positive and negative radio-mechanical feedback, where the radio jets are heating the hot atmosphere of the BCG and driving gas flows while triggering star formation in the outflowing gas. Although determining the velocities of young stellar populations in the star forming filaments is unfeasible in this case, the stars formed in outflowing gas are expected to have radial orbits and they could populate outer regions of galaxies \citep{ishibashi12}. Positive feedback is expected to be more common at high redshift when gas densities were higher and AGN feedback was dominant. Some models predict that star formation triggered by positive feedback contributes substantially to star formation rates in distant galaxies \citep{silk13} and it may be necessary to explain the $M_{\rm BH}-\sigma$ relation observed in galaxies \citep{silk10}. This object represents an excellent example of star formation at the interface of radio jet/gas interaction and is helpful in understanding the positive feedback of the radio jets in galaxies in the distant Universe.

\section*{Acknowledgements}

The authors gratefully acknowledge the anonymous referee for their comments, which helped us to improve the paper. BRM acknowledges support from the Natural Sciences and Engineering Council of Canada and the Canadian Space Agency Space Science Enhancement Program. HRR acknowledges support from an STFC Ernest Rutherford Fellowship and an Anne McLaren Fellowship. ACE acknowledges support from STFC grant ST/P00541/1. We thank Laura Birzan for providing VLA radio maps of Abell 1795. Based on observations collected at the European Southern Observatory under ESO programme(s) 60.A-9022(C) and 094.A-0859(A). This paper makes use of the following ALMA data: ADS/JAO.ALMA\#2015.1.00623.S. ALMA is a partnership of ESO (representing its member states), NSF (USA) and NINS (Japan), together with NRC (Canada), NSC and ASIAA (Taiwan), and KASI (Republic of Korea), in cooperation with the Republic of Chile. The Joint ALMA Observatory is operated by ESO, AUI/NRAO and NAOJ. Based on observations made with the NASA/ESA Hubble Space Telescope, and obtained from the Hubble Legacy Archive, which is a collaboration between the Space Telescope Science Institute (STScI/NASA), the Space Telescope European Coordinating Facility (ST-ECF/ESA) and the Canadian Astronomy Data Centre (CADC/NRC/CSA).

This research made use of {\sc python} \citep{rossum09}, {\sc Astropy} \citep{astropy13,astropy18}, {\sc matplotlib} \citep{hunter07}, {\sc numpy} \citep{walt11,harris20}, and {\sc scipy} \citep{scipy20}. We thank their developers for maintaining them and making them freely available.

\section*{Data Availability}

All data used in this paper is publicly available from the respective telescope's data archive website. Processed data products will be made available upon a reasonable request to the corresponding author.



\bibliographystyle{mnras}
\bibliography{ref} 




\appendix

\section{Additional figures}
The appendix contains additional figures showing the H$\alpha$ and [NII] emission lines in the XSHOOTER spectrum of the star forming filament in Abell 1795 referenced in section~\ref{xshoo} and the velocity maps of the molecular gas and H$\alpha$ gas showing the regions used to calculate the VSF referenced in section~\ref{vsf}.

\begin{figure*}
    \centering
    \includegraphics[width=1.1\textwidth]{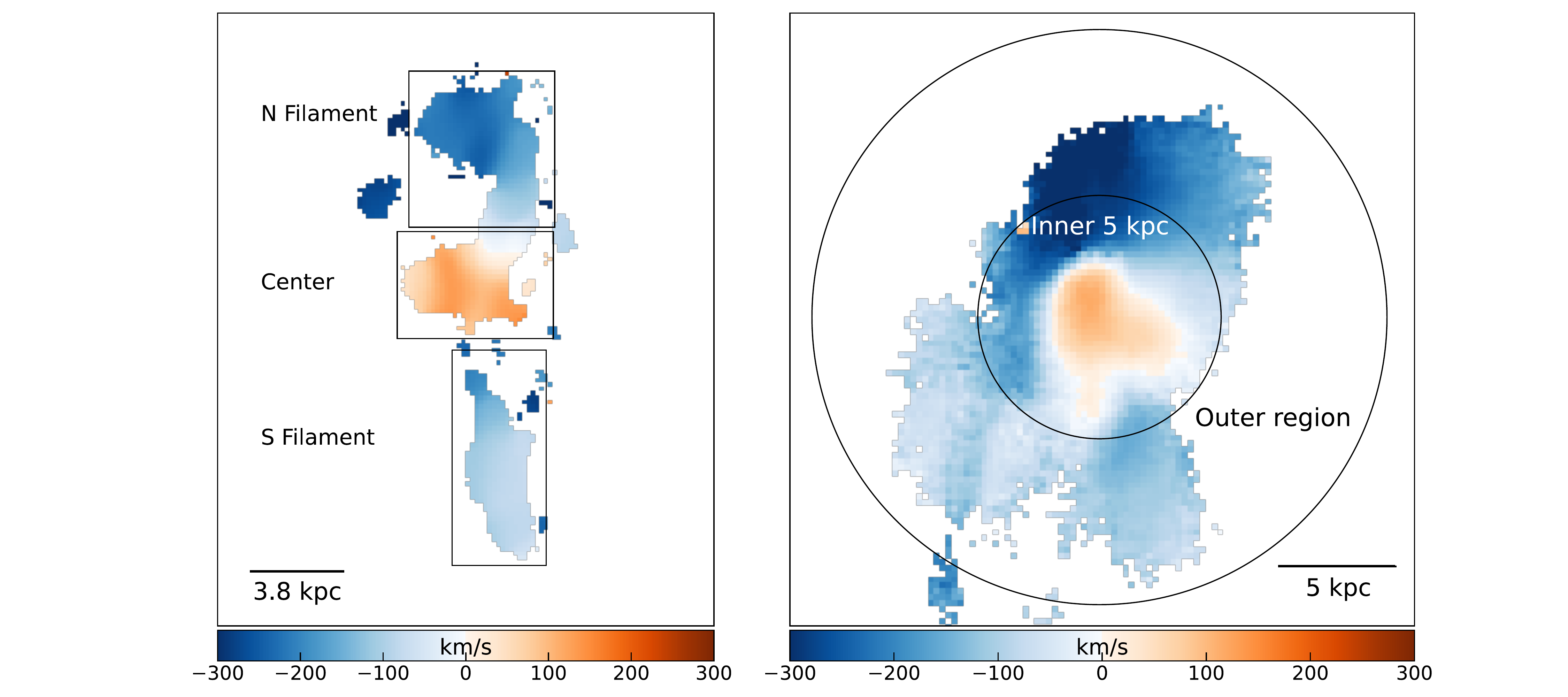}
    \caption{The velocity maps of the molecular gas (left panel) and H$\alpha$ gas (right panel) are shown with respective regions used to calculate the velocity structure function as described in section~\ref{vsf}, where the error in velocity at each pixel is less than 10 km s$^{-1}$.}
    \label{fig:vsfmaps}
\end{figure*}

\begin{figure*}
    \centering
    \includegraphics[width=\textwidth]{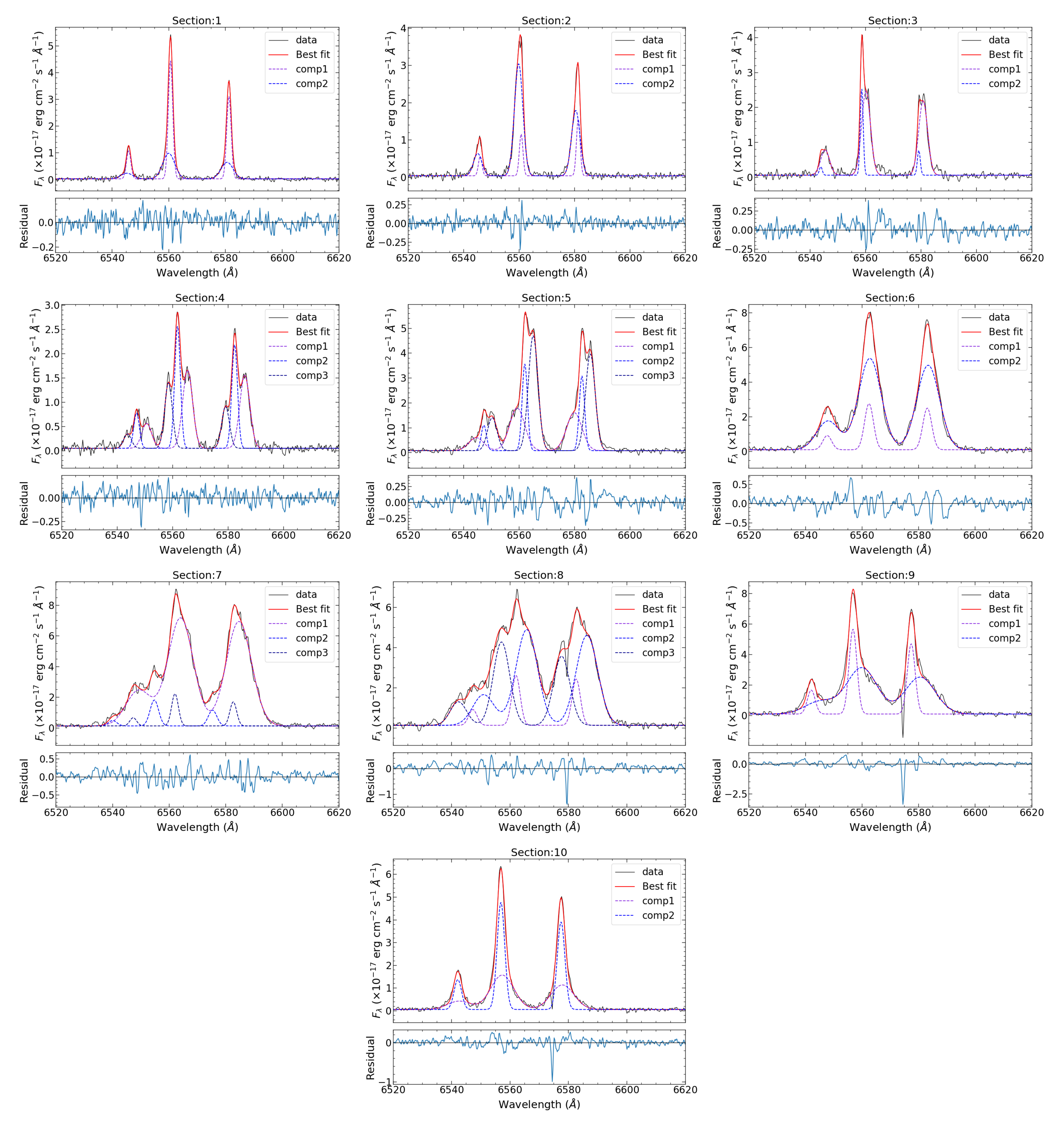}
    \caption{H$\alpha$-[NII] complex emission line fits to the regions along the slits as described in section~\ref{ionizedgas}. The red curve is the summed emission line profile and the blue curves show individual gaussian components required for the fit. The residuals are shown in the bottom panel of each fit.}
    \label{fig:Hafits}
\end{figure*}


\bsp	
\label{lastpage}
\end{document}